\documentclass[12pt,preprint]{aastex}
\usepackage{subfigure}

\newcommand{\bm}[1]{\mbox{\boldmath{$#1$}}}

\newcommand{\del}{{\bf \nabla}}
\newcommand{\alf}{{\rm Alfv\acute{e}n}}
\newcommand{\pr}{P_{\rm m}}
\newcommand{\cs}{c_{\rm s}}
\newcommand{\va}{v_{\rm A}}
\newcommand{\pavg}{\langle\langle P\rangle\rangle}
\newcommand{\maxwell}{\langle\langle-B_xB_y\rangle\rangle}
\newcommand{\reynolds}{\langle\langle \rho v_x\delta v_y\rangle\rangle}
\newcommand{\magnetic}{\langle\langle B^2/2 \rangle\rangle}
\newcommand{\magx}{\langle\langle B_x^2/2 \rangle\rangle}
\newcommand{\magy}{\langle\langle B_y^2/2 \rangle\rangle}
\newcommand{\magz}{\langle\langle B_z^2/2 \rangle\rangle}

\newcommand{\pertkin}{\langle\langle \rho \delta v^2/2\rangle\rangle}
\newcommand{\kinx}{\langle\langle \rho v_x^2/2\rangle\rangle}
\newcommand{\pertkiny}{\langle\langle \rho \delta v_y^2/2\rangle\rangle}
\newcommand{\kinz}{\langle\langle \rho v_z^2/2\rangle\rangle}

\begin{document}

\title{Resistivity-driven State Changes in Vertically Stratified Accretion Disks}

\author{Jacob B. Simon\textsuperscript{1}, John F. Hawley\textsuperscript{2}, Kris Beckwith\textsuperscript{3}}

\begin{abstract}

We investigate the effect of shear viscosity, $\nu$, and Ohmic resistivity, $\eta$ on the magnetorotational instability (MRI) in vertically stratified accretion disks through a series of local simulations with the {\it Athena} code. First, we use a series of unstratified simulations to calibrate physical dissipation as a function of resolution and background field strength; the effect of the magnetic Prandtl number, $\pr = \nu/\eta$, on the turbulence is captured by $\sim 32$ grid zones per disk scale height, $H$. In agreement with previous results, our stratified disk calculations are characterized by a subthermal, predominately toroidal magnetic field that produces MRI-driven turbulence for $|z| \lesssim 2 H$. Above $|z| \sim 2 H$, magnetic pressure dominates and the field is buoyantly unstable.  Large scale radial and toroidal fields are also generated near the mid-plane and subsequently rise through the disk. The polarity of this mean field switches on a roughly 10 orbit period in a process that is well-modeled by an $\alpha$--$\Omega$ dynamo. Turbulent stress increases with $\pr$ but with a shallower dependence compared to unstratified simulations. For sufficiently large resistivity, $\eta \sim \cs H/1000$ where $\cs$ is the sound speed, MRI turbulence within $2 H$ of the mid-plane undergoes periods of resistive decay followed by regrowth. This regrowth is caused by amplification of toroidal field via the dynamo. This process results in large amplitude variability in the stress on 10 to 100 orbital timescales, which may have relevance for partially ionized disks that are observed to have high and low accretion states.

\end{abstract}


\noindent
{\footnotesize \textsuperscript{1} jbsimon@jila.colorado.edu, JILA, University of Colorado, 440 UCB, Boulder, CO 80309-0440} \\
\indent{\footnotesize Department of Astronomy, University of Virginia, P.O. Box 400325, Charlottesville, VA 22904-4325} \\
\indent{\footnotesize \textsuperscript{2}  jh8h@virginia.edu, Department of Astronomy, University of Virginia, P.O. Box 400325} \\
\indent{\footnotesize Charlottesville, VA 22904-4325} \\
\indent{\footnotesize \textsuperscript{3} kris.beckwith@jila.colorado.edu, JILA, University of Colorado, 440 UCB, Boulder, CO 80309-0440}
% SECTION 1

\section{Introduction}

Disk accretion is a fundamental astrophysical process, responsible
for such diverse phenomena as the immensely luminous, distant quasars and the
formation and evolution of protostellar systems.  This process requires
removing angular momentum from the orbiting gas, and from the beginnings of
disk theory it has been clear that microphysical viscosity is orders
of magnitude too small to account for observed accretion rates.  It is
now understood that accretion is driven by magnetohydrodynamic (MHD)
turbulent stresses arising from a robust and powerful instability known
as the magnetorotational instability (MRI) \cite[][]{balbus91,balbus98}.

Analytic investigations of the MRI have proven to be very insightful
\cite[e.g.,][]{balbus91,goodman94}, but they can offer only limited
guidance to the behavior of the MRI in the fully turbulent saturated
state.  Numerical simulations have become the essential tool for
investigating how the MRI operates in accretion disks, and in particular, local shearing box
simulations \cite[see][]{hawley95a} have proven to be especially useful in such
studies.  The shearing box system solves the equations of
MHD in a local, co-rotating patch of an accretion disk. The size of this
patch is assumed to be small compared to its radial distance from the
central star, allowing one to use Cartesian geometry while retaining the
essential dynamics of differential rotation.  Shearing box simulations
have led to an improved understanding of MRI turbulence while addressing
some basic questions about accretion disks.

One such question is, what sets the saturation level of MRI-driven
turbulence and the angular momentum transport rate?  This is usually
phrased as ``What is $\alpha$?'' following \cite{shakura73} who
made the ansatz that the $r \phi$ component of the turbulent stress,
$\tau_{r \phi}$, is proportional to the pressure, $\tau_{r \phi}=\alpha P$.
Previous shearing box simulations have found substantial evidence that MRI-driven
stress does not behave in a manner consistent with how $\alpha$ is
often applied in disk theory.  The early studies of \cite{hawley95a}
showed that the stress is proportional to the {\it magnetic} pressure,
but the magnetic pressure is not itself directly determined by the
gas pressure, a result that holds over a wide range of shearing box
simulations \citep{blackman08}.  \cite{sano04} performed an extensive
survey and observed, at best, only a very weak gas pressure dependence for
turbulent stress.  More recently, vertically stratified local simulations
of radiation-dominated disks \cite[][]{hirose09a} have shown that while there
is a correlation between the MRI stress and total (gas plus radiation)
pressure, it has the opposite causal relationship from that normally
assumed in the $\alpha$ model:  the stress determines the pressure, not
the other way around.  Increased stress can lead to increased pressure
through turbulent heating, but an increase in pressure does not feed
back into the stress.

What then does determine the saturation level of the MRI?  We know that
viscosity and resistivity can significantly affect the properties of the
linear MRI by reducing growth rates and altering the stability limits.
Recently, the influence of the viscous and Ohmic dissipation on MRI-induced
turbulence has become a focus of shearing box simulations.  Previous
investigations into the effect of Ohmic resistivity on the saturated state
\cite[]{hawley96,sano98,fleming00,sano01,ziegler01,sano02b} found that
angular momentum transport decreases as the resistivity is increased,
but it was not until the very recent work of \cite{fromang07a},
\cite{pessah07}, and \cite{fromang07b} that the influence of
physical dissipation in numerical simulations was fully appreciated.
\cite{fromang07a} and \cite{pessah07} found that for shearing box
simulations without a net magnetic flux and without explicit dissipation
terms, the turbulent saturation level decreases with increasing
numerical resolution without any sign of convergence. This surprising
result was subsequently confirmed with different numerical codes
\cite[e.g.,][]{simon09a,guan09a}.  In a second paper, \cite{fromang07b}
demonstrated that the saturation level of the MRI in shearing boxes depends
strongly on the magnetic Prandtl number, i.e., the ratio of viscosity
to resistivity ($\pr = \nu/\eta$).  Specifically, for local simulations
without a net magnetic flux, the turbulent stresses increase nearly
linearly with $\pr$ and for $\pr \lesssim 1$, the turbulence decays.
\cite{fromang10} subsequently showed that at a fixed $\pr$, the presence
of even a small viscosity and resistivity is sufficient to provide
convergence in the zero-net field case.  The increase in stress with $\pr$
is also present for net vertical fields \cite[]{lesur07} and net toroidal
fields \cite[]{simon09b}; for these field geometries, the $\pr$
dependence is weaker and turbulence is maintained even for $\pr < 1$.
For the net toroidal field model, only a sufficiently high resistivity
could actually kill the turbulence completely \cite[]{simon09b}.

Most of these viscosity and resistivity studies used the unstratified
shearing box model where vertical gravity is ignored and periodic
boundary conditions are assumed for the vertical direction.  Shearing box
models that include vertical stratification are more realistic however,
and interesting new effects arise when vertical gravity is included.
For example, in most such simulations, net toroidal field is
generated near the mid-plane via MRI turbulence, buoyantly rises upwards,
and is replaced with a field of the opposite sign in the mid-plane
region.  This behavior happens on a timescale of $\sim 10$ orbits
and appears to be indicative of an MHD dynamo operating within the disk
\cite[e.g.,][]{brandenburg95,stone96a,hirose06,guan10,shi10,gressel10,davis10}.
Furthermore, the vertical structure of the disk consists of MRI-turbulent
gas that is marginally stable to buoyancy within $|z| \sim 2 H$, whereas
outside of this region, the gas is magnetically dominated, significantly
less turbulent, and buoyantly unstable \cite[e.g.,][]{guan10,shi10}.

The effects of resistivity on the MRI in vertically stratified
shearing boxes has been studied \citep[e.g.,][]{miller00},
but many of these calculations are designed with protostellar
systems in mind and employ a very large resistivity to completely
quench the MRI  and create a ``dead zone'' near the midplane
\cite[e.g.,][]{gammie96,fleming03,fromang06a,oishi07,turner08,ilgner08,oishi09,turner10}.
The effects of smaller resistivities and of $\nu$ and $\pr$ on vertically
stratified turbulence has barely been examined.  Recently, however,
\cite{davis10} investigated how the results of \cite{fromang07a} and
\cite{fromang07b} might change in the presence of vertical stratification.
\cite{davis10} considered the zero-net magnetic flux case and employed
vertically periodic boundary conditions to ensure that zero net
flux was maintained throughout the simulation.  They found that without
physical dissipation, the volume-averaged stress level reaches a constant
value as numerical resolution is increased, in contrast to unstratified
simulations where the stress declines.  They further examined three $\nu$
and $\eta$ values that lead to a decay of turbulence in unstratified
boxes. With vertical gravity, however, these $\nu$ and $\eta$ values lead
to large, long timescale fluctuations in the volume averaged stress level;
the turbulence saturates at a relatively high level for $\sim$ 100 orbits,
then decreases to a lower saturation level for another $\sim$ 100 orbits,
and then increases again.

The primary goal of this work is to extend these first results and
investigate in more detail how $\nu$ and $\eta$ affect MRI turbulence
in vertically stratified shearing boxes.  In particular, we consider
the origin of the long-term fluctuations seen in \cite{davis10},
and whether that effect might be relevant to real accretion disks.
Since stratification can alter the saturation levels of MRI-induced
turbulence, we also investigate whether increasing $\pr$ still leads to
increased stress levels and how $\pr$ affects the dynamo
process previously observed in stratified simulations.  Finally, these
simulations will also serve as an essential starting point for future
studies that include more realistic physics, such as temperature-
and density-dependent $\nu$ and $\eta$.

The structure of this paper is as follows.  In \S~\ref{method},
we describe our evolution equations, parameters, and initial conditions.  In
\S~\ref{unstrat_sims}, we present a series of unstratified shearing box
simulations to calibrate the effects of physical dissipation and serve
as controls for the vertically stratified shearing boxes with constant
$\nu$ and $\eta$.  We discuss our vertically stratified simulations
in \S~\ref{strat_sims}, which are the primary focus of this paper.
The first set of these simulations contain no physical dissipation,
and we carry out several analyses to improve our understanding of
vertically stratified MRI turbulence.  The second set of simulations
then includes physical dissipation to study the $\pr$ effect.  We wrap
up with a discussion and our general conclusions in \S~\ref{discussion}.
We also present a detailed description of our numerical algorithm
in an Appendix.

% SECTION 2

\section{Method}
\label{method}

In this study, we use {\it Athena}, a second-order accurate Godunov
flux-conservative code for solving the equations of MHD.  \textit{Athena}
uses the dimensionally unsplit corner transport upwind (CTU) method
of \cite{colella90} coupled with the third-order in space piecewise
parabolic method (PPM) of \cite{colella84} and a constrained transport
\citep[CT;][]{evans88} algorithm for preserving the $\del \cdot {\bm
B}$~=~0 constraint.  We use the HLLD Riemann solver to calculate the
numerical fluxes \cite[]{miyoshi05,mignone07b}.  A detailed description
of the \textit{Athena} algorithm and the results of various test problems
are given in \cite{gardiner05a}, \cite{gardiner08}, and \cite{stone08}.

Our simulations utilize the shearing box approximation, a model
for a local co-rotating disk patch whose size is small compared to the
radial distance from the central object, $R_o$.  We construct a local
Cartesian frame, $x=(R-R_o)$, $y=R_o \phi$, and $z$,
co-rotating with an angular velocity $\Omega$ corresponding to
the orbital frequency at $R_o$, the center of the box.  In this 
frame, the equations of motion become \citep{hawley95a}:

\begin{equation}
\label{continuity_eqn}
\frac{\partial \rho}{\partial t} + \del \cdot (\rho {\bm v}) = 0,
\end{equation}
\begin{equation}
\label{momentum_eqn}
\frac{\partial \rho {\bm v}}{\partial t} + \del \cdot \left(\rho {\bm v}{\bm v} - {\bm B}{\bm B}\right) + \del \left(P + \frac{1}{2} B^2\right) = 2 q \rho \Omega^2 {\bm x} - \rho \Omega^2 {\bm z} - 2 {\bm \Omega} \times \rho {\bm v} + \del \cdot {\bm T},
\end{equation}
\begin{equation}
\label{induction_eqn}
\frac{\partial {\bm B}}{\partial t} - \del \times \left({\bm v} \times {\bm B}\right) =  -\del \times \left(\eta \del \times {\bm B}\right).
\end{equation} 

\noindent 
where $\rho$ is the mass density, $\rho {\bm v}$ is the momentum
density, ${\bm B}$ is the magnetic field, $P$ is the gas pressure,
and $q$ is the shear parameter, defined as $q = -d$ln$\Omega/d$ln$R$.
We use $q = 3/2$, appropriate for a Keplerian disk.  We
assume an isothermal equation of state $P = \rho \cs^2$, where $\cs$
is the isothermal sound speed.  From left to right, the source terms
in equation~(\ref{momentum_eqn}) correspond to radial tidal forces
(gravity and centrifugal), vertical gravity, the Coriolis force, and
the divergence of the viscous stress tensor, ${\bm T}$, defined as

\begin{equation} 
\label{viscous_stress_tensor} 
T_{ij} = \rho \nu
\left(\frac{\partial v_i}{\partial x_j} + \frac{\partial v_j}{\partial
x_i} - \frac{2}{3}\delta_{ij} \del \cdot {\bm v} \right), 
\end{equation}

\noindent where the indices refer to the spatial components
\cite[]{landau59}, and $\nu$ is the shear viscosity.  We neglect bulk
viscosity.  The source term in equation~(\ref{induction_eqn}) is the
effect of Ohmic resistivity, $\eta$, on the magnetic field evolution.
Note that our system of units has the magnetic permeability $\mu = 1$.
Details about the numerical integration of these equations are presented in
the Appendix.

For unstratified shearing box simulations, the boundary conditions are
periodic for $y$ and $z$, and shearing-periodic for $x$.  In
stratified simulations, the periodic $z$ boundary conditions are 
replaced with outflow boundary conditions.  The specifics are
described in the Appendix.

\subsection{Dissipation Parameters}

In all of our simulations, $\nu$ and $\eta$ are parameterized in terms of
Reynolds numbers.  At a scale height, $H$, away from the radial center
of the shearing box, the fluid velocity is $|v_y| \sim q H \Omega \sim \cs$.  Thus,
the sound speed is a representative velocity for the fluid, and 
we define the Reynolds numbers as

\begin{equation}
\label{reynolds}
Re \equiv \frac{\cs H }{\nu}.
\end{equation}

\noindent
Similarly we define the magnetic Reynolds number,

\begin{equation}
\label{mag_reynolds}
Rm \equiv \frac{\cs H}{\eta},
\end{equation}

\noindent
and their ratio, the magnetic Prandtl number,

\begin{equation}
\label{prandtl}
\pr \equiv \frac{\nu}{\eta} = \frac{Rm}{Re}.
\end{equation}

Because resistivity affects the MRI directly, another useful dimensionless
quantity is the Elsasser number

\begin{equation}
\label{elsasser}
\Lambda \equiv \frac{\va^2}{\eta \Omega}.
\end{equation}

\noindent 
$\Lambda$ can be computed on a zone-by-zone basis, but it is also helpful to
calculate an average $\Lambda$ for a simulation.  
Thus, we can write the characteristic $\alf$ speed in direction $i$
via the averaged magnetic field in that direction,

\begin{equation}
\label{alfven_q}
v_{{\rm A}i} = \sqrt{\frac{\langle B_i^2\rangle}{\langle\rho\rangle}},
\end{equation}

\noindent where the angled brackets denote a volume average, and
the subscript $i = (x,y,z)$ depending on the direction of interest.  
In the above definition of $\Lambda$, all three components are used in
calculating the $\alf$ speed. 
As the Elsasser number approaches unity, resistivity becomes more
dominant, stabilizing the MRI.   \cite{simon09b} found
that turbulence decayed in net toroidal field, unstratified shearing boxes
when the Elsasser number was $\lesssim 100$.

Although we focus on physical dissipation, there is numerical
dissipation as well.  One way to characterize the effects of finite
resolution is to compare the characteristic MRI wavelength,
$\lambda_{\rm MRI} = 2\pi v_{\rm A} /\Omega$ to the grid zone size.
\cite{noble10} defined this ``quality factor'' $Q$ as

\begin{equation}
\label{q_resolve}
Q_i \equiv \frac{\lambda_{{\rm MRI},i}}{\Delta x_i} = \frac{2 \pi
v_{{\rm A}i}}{\Omega \Delta x_i}
\end{equation}

\noindent
Again, $v_{{\rm A}i}$ is defined via equation~(\ref{alfven_q}).
For $Q_i \lesssim 6$, the growth of the MRI can be reduced
\cite[]{sano04} and the MRI is under-resolved, although
this number has some uncertainty and should be taken only as an estimate.

One can also calculate a volume-averaged $\alf$
speed rather than an $\alf$ speed determined by the averaged field,
i.e.,
\begin{equation}
\label{alfven_q_alternate}
v_{{\rm A}i} = \left\langle \frac{|B_i|}{\sqrt{\rho}}\right\rangle.
\end{equation}

\noindent For convenience, most of our analysis uses
equation~(\ref{alfven_q}), as the volume-averaged history data and the one-dimensional, horizontally averaged
quantities are routinely computed at high time
resolution for use in creating, e.g., space-time plots. 
As a check, we have calculated the volume-averaged $\alf$ speed via equation~(\ref{alfven_q_alternate})
for several hundreds of orbits in a few of our vertically stratified simulations (the volume average is done within
$2 H$ of the mid-plane).  We found that $\va$ from equation~(\ref{alfven_q_alternate}) is roughly 1-1.5 times
larger than that from equation~(\ref{alfven_q}).

\subsection{Parameters and Initial Conditions}
\label{parameters}

We have run shearing box simulations with different field geometries,
dissipation values and resolutions, both with and without vertical
gravity.  Here, we describe the parameters and initial conditions used in
these two types of simulations.  The results from the unstratified simulations
are presented in \S~\ref{unstrat_sims} and the results from the stratified
simulations are in \S~\ref{strat_sims}.

\subsubsection{Sine Z Unstratified Simulations}
\label{sinez}

The first set of simulations are unstratified, zero net magnetic flux
shearing boxes.  This is the same type of problem as investigated 
in \cite{fromang07b} and \cite{simon09b}.  The initial magnetic field is 
\begin{equation} 
{\bm B} = \sqrt{2 P_o/\beta} ~ {\rm sin}(2 \pi x/L_x) ~ \hat{\bm z} 
\end{equation}
with $\beta = 400$. The isothermal sound speed is $\cs = 0.001$,
corresponding to an initial gas pressure $P_o = 10^{-6}$ with initial
density $\rho_o = 1$.  The orbital velocity of the local domain is
$\Omega = 0.001$.  For these simulations, we define the scale height to be 
\begin{equation}
\label{h_sz}
H \equiv \frac{\cs}{\Omega} = 1, 
\end{equation}

\noindent which is a slightly different definition than that in the
stratified box (by a factor of $\sqrt{2}$).  The size of the box is
$L_x = 1 H$, $L_y = 4 H$, and $L_z = 1 H$. We run several resolutions
in order to study convergence, and at each resolution, we study four different cases of $Rm$ and $\pr$
values. See Table~\ref{tbl:unstrat_sims} for the parameters of these
runs.  The labeling scheme of the
runs refers to resolution, field geometry, and dissipation values;
e.g.,16SZRm12800Pm16 corresponds to 16 zones per $H$, ``SZ" for the
``Sine Z" geometry, and $Rm = 12800$, $\pr = 16$.  Since, as we will see,
$Rm$ is the critical parameter in many of our simulations, we choose to
include it as part of the run label, differing from the convention of
related works \cite[e.g.,][]{fromang07b,simon09b}.

The MRI is seeded with random perturbations to the density and the
velocity components introduced at the grid scale.  The amplitude of the
density perturbations is $\delta \rho = 0.01$ and the amplitude of the
velocity perturbations is $(1/5) \delta \rho \cs$ for each component
(a different perturbation is applied for each component).  We do not
employ orbital advection in these simulations (see Appendix).   All simulations are run
to 400 orbits, except for the runs in which the turbulence decays and
also 32SZRm12800Pm16 and 32SZRe12500Pm4, which were run to 289 orbits
and 246 orbits, respectively. We also include the results from previous,
higher resolution simulations \citep{simon09b} in our analysis.

\subsubsection{Flux Tube Unstratified Simulations}
\label{flux_tube}

The second set of unstratified simulations contain a net toroidal field
and are initialized with the
twisted azimuthal flux tube of \cite{hirose06}, with minor modifications
to the dimensions and $\beta$ values.
The initial conditions are designed to match those used
in the vertically stratified simulations of \S~\ref{vert_strat_init},
so we define $H$ to be that of a stratified, isothermal disk,

\begin{equation}
\label{scale_height}
H = \frac{\sqrt{2} \cs}{\Omega}.
\end{equation}

\noindent The isothermal sound speed, $\cs = 7.07 \times 10^{-4}$,
corresponding to an initial value for the gas pressure of $P_o = 5
\times 10^{-7}$.  With $\Omega = 0.001$, the value for the scale height
is $H = 1$.  As in the SZ simulations, random
perturbations are added to the density and velocity components.

The initial toroidal field, $B_y$, is given by
\begin{equation}
\label{toroidal}
B_y = \left\{ \begin{array}{ll}
\sqrt{\frac{2 P_o}{\beta_y} - \left(B_x^2+B_z^2\right)}& \quad
\mbox{if $B_x^2 + B_z^2 \neq 0$} \\
0 & \quad 
\mbox{if $B_x^2 + B_z^2 = 0$}
\end{array} \right.
\end{equation}

\noindent and we have run simulations with a toroidal field $\beta$
of $\beta_y =$ 100, 1000, and 10000.  
The poloidal field components, $B_x$ and $B_z$,
are calculated from the $y$ component of the vector potential,

\begin{equation}
\label{vector_potential}
A_y = \left\{ \begin{array}{ll}
- \sqrt{\frac{2 P_o}{\beta_p}}\frac{H}{2 \pi} \left[1 + {\rm cos}\left(\frac{2 \pi r}{H}\right)\right] & \quad
\mbox{if $r < \frac{H}{2}$} \\
0 & \quad 
\mbox{if $r \geq \frac{H}{2}$}
\end{array} \right.
\end{equation}
\noindent where $r = \sqrt{x^2 + z^2}$ and $\beta_p = 1600$ is the
poloidal field $\beta$ value.

The labeling convention is the same as in \S~\ref{sinez}, but with ``FT"
for ``Flux Tube" (see Table~\ref{tbl:unstrat_sims}).  We run simulations
both with and without physical dissipation.  Those simulations with no
physical dissipation are labeled with ``Num" for ``Numerical dissipation",
and for the $\beta_y = 1000$ (10000) cases, we append $\beta1000$
($\beta10000$) on the end of the run label.  The domain size is $L_x =
2 H$, $L_y = 4 H$, and $L_z = 1 H$, and the resolution is 32 zones per
$H$. Orbital advection is employed in these calculations (see Appendix).  32FTNum was
run to 150 orbits, and both 32FTNum$\beta1000$ and 32FTNum$\beta10000$
were run to 110 orbits.

The runs with physical dissipation are initiated from the turbulent state
(at $t = 100$ orbits) of the corresponding $\beta_y$ value run with only
numerical dissipation.  These simulations were all run out to 220 orbits.

\begin{deluxetable}{l|cccccc}
\tabletypesize{\scriptsize}
\tablewidth{0pc}
\tablecaption{Unstratified Simulations\label{tbl:unstrat_sims}}
\tablehead{
\colhead{Label}&
\colhead{$Re$}&
\colhead{$Rm$}&
\colhead{$\pr$}&
\colhead{Resolution}&
\colhead{$\alpha$} &
\colhead{Description} \\
\colhead{ }&
\colhead{ }&
\colhead{ }&
\colhead{ }&
\colhead{(zones per $H$)}&
\colhead{ }&
\colhead{ }    } 
\startdata
16SZRm12800Pm16 & 800 & 12800 & 16 & 16 & 0.011 & zero net flux \\
32SZRm12800Pm16 & 800 & 12800 & 16 & 32 & 0.033 & zero net flux \\
64SZRm12800Pm16 & 800 & 12800 & 16 & 64 & 0.042 & zero net flux \\
128SZRm12800Pm16\tablenotemark{a} & 800 & 12800 & 16 & 128 & 0.046 & zero net flux \\
16SZRm12500Pm4 & 3125 & 12500 & 4 & 16 & 0.0043 & zero net flux \\
32SZRm12500Pm4 & 3125 & 12500 & 4 & 32 & 0.013 & zero net flux \\
64SZRm12500Pm4 & 3125 & 12500 & 4 & 64 & 0.015 & zero net flux \\
128SZRm12500Pm4\tablenotemark{a} & 3125 & 12500 & 4 & 128 & 0.013 & zero net flux \\
16SZRm6250Pm1 & 6250 & 6250 & 1 & 16 & decay & zero net flux \\
32SZRm6250Pm1 & 6250 & 6250 & 1 & 32 & decay & zero net flux \\
64SZRm6250Pm1 & 6250 & 6250 & 1 & 64 & decay & zero net flux \\
16SZRm25600Pm2 & 12800 & 25600 & 2 & 16 & 0.0077 & zero net flux \\
32SZRm25600Pm2 & 12800 & 25600 & 2 & 32 & 0.010 & zero net flux \\
64SZRm25600Pm2 & 12800 & 25600 & 2 & 64 & 0.0078 & zero net flux \\
32FTNum & -- & -- & -- & 32 & 0.021\tablenotemark{b} & flux tube, num. dissipation, $\beta = 100$\\
32FTNum$\beta$1000 & -- & -- & -- & 32 & 0.020\tablenotemark{c} & flux tube, num. dissipation, $\beta = 1000$ \\
32FTNum$\beta$10000 & -- & -- & -- & 32 & 0.018\tablenotemark{c} & flux tube, num. dissipation, $\beta = 10000$ \\
32FTRm800Pm0.5 & 1600 & 800 & 0.5 & 32 & decay & restarted from 32FTNum \\
32FTRm3200Pm0.5 & 6400 & 3200 & 0.5 & 32 & 0.0094 & restarted from 32FTNum \\
32FTRm3200Pm2 & 1600 & 3200 & 2 & 32 & 0.018 & restarted from 32FTNum \\
32FTRm3200Pm4 & 800 & 3200 & 4 & 32 & 0.028 & restarted from 32FTNum \\
32FTRm6400Pm4 & 1600 & 6400 & 4 & 32 & 0.029 & restarted from 32FTNum \\
32FTRm6400Pm8 & 800 & 6400 & 8 & 32 & 0.041 & restarted from 32FTNum \\
32FTRm1600Pm1$\beta$1000 & 1600 & 1600 & 1 & 32 & decay & restarted from 32FTNum$\beta$1000 \\
32FTRm3200Pm2$\beta$1000 & 1600 & 3200 & 2 & 32 & 0.015\tablenotemark{d} & restarted from 32FTNum$\beta$1000 \\
32FTRm3200Pm2$\beta$10000 & 1600 & 3200 & 2 & 32 & decay & restarted from 32FTNum$\beta$10000 \\
\enddata
\tablenotetext{a}{These runs were taken from \cite{simon09b}}
\tablenotetext{b}{Time averaged from orbit 20 to 150}
\tablenotetext{c}{Time averaged from orbit 20 to 110}
\tablenotetext{d}{Time averaged from orbit 120 to 400}
\end{deluxetable}

\subsubsection{Vertically Stratified Simulations}
\label{vert_strat_init}

In vertically stratified shearing boxes, the initial density corresponds to
an isothermal hydrostatic equilibrium

\begin{equation}
\label{density_init}
\rho(x,y,z) = \rho_o {\rm exp}\left(-\frac{z^2}{H^2}\right),
\end{equation}

\noindent where $\rho_o = 1$ is the mid-plane density, and $H$
is the scale height in the disk, as defined in \S~\ref{flux_tube}.
A density floor of $10^{-4}$ is applied to the physical domain as
too small a density leads to a large $\alf$ speed and a very small
timestep.  Furthermore, numerical errors in energy make it difficult
to evolve regions of very small plasma $\beta$.  All other parameters
and initial conditions are identical to the corresponding FT runs of
\S~\ref{flux_tube}.  All the stratified runs use $\beta_y = 100$ for
the initial toroidal field strength.

The domain size for all vertically stratified simulations is $L_x =
2 H$, $L_y = 4 H$, and $L_z = 8 H$, and we have run most simulations
at 32 grid zones per $H$, with a few at 64 zones per $H$.  Orbital
advection is employed in all of these runs (see Appendix).  The runs are listed in
Table~\ref{tbl:strat_sims}.  The labeling scheme is the same as for
the unstratified simulations, but we omit the ``FT'' label as all the
stratified runs use the flux tube initial conditions. 
We study the effect of physical dissipation by restarting the equivalent
resolution, numerical-dissipation-only run at $t = 100$ orbits. 

\begin{deluxetable}{l|cccccc}
\tabletypesize{\scriptsize}
\tablewidth{0pc}
\tablecaption{Vertically Stratified Simulations\label{tbl:strat_sims}}
\tablehead{
\colhead{Label}&
\colhead{$Re$}&
\colhead{$Rm$}&
\colhead{$\pr$}&
\colhead{Resolution}&
\colhead{Integration time}&
\colhead{Description} \\
\colhead{ }&
\colhead{ }&
\colhead{ }&
\colhead{ }&
\colhead{(zones per $H$)}&
\colhead{$t_{\rm stop} - t_{\rm start}$ (orbits)}&
\colhead{ }    } 
\startdata
32Num & -- & -- & -- & 32 & 1058 & num. dissipation \\
32Rm800Pm0.5 & 1600 & 800 & 0.5 & 32 & 263 & -- \\
32Rm3200Pm0.5 & 6400 & 3200 & 0.5 & 32 & 863 & --\\
32Rm3200Pm2 & 1600 & 3200 & 2 & 32 & 1082 & -- \\
32Rm3200Pm4 & 800 & 3200 & 4 & 32 & 487 & -- \\
32Rm6250Pm1 & 6250 & 6250 & 1 & 32 & 337 & -- \\
32Rm6400Pm8 & 800 & 6400 & 8 & 32 & 325 & -- \\
32Rm6400Pm4 & 1600 & 6400 & 4 & 32 & 488 & -- \\
32Rm3200Pm2\_By+ & 1600 & 3200 & 2 & 32 & 584 & $B_y$ added at 150 orbits \\
32ShearBx & -- & -- & -- & 32 & 45 & net $B_x$ within midplane \\
64Num & -- & -- & -- & 64 & 159 & num. dissipation \\
64Rm800Pm0.5 & 1600 & 800 & 0.5 & 64 & 108 & -- \\
64Rm3200Pm0.5 & 6400 & 3200 & 0.5 & 64 & 83 & -- \\
64Rm3200Pm2 & 1600 & 3200 & 2 & 64 & 80 & -- \\
64Rm3200Pm4 & 800 & 3200 & 4 & 64 & 84 & -- \\
64Rm6400Pm4 & 1600 & 6400 & 4 & 64 & 80 & -- \\
\enddata
\end{deluxetable}

% SECTION 3

\section{Calibration of Physical Dissipation in Unstratified Disks}
\label{unstrat_sims}

While our focus is on the effects of physical dissipation on MRI-driven
turbulence in stratified disks, we have carried out a series of unstratified shearing
box simulations to address several points.  First, what resolution is
needed to capture the influence of physical dissipation, and how does a
given resolution influence the effects due to dissipation terms? Second, what is
the effect of physical dissipation on different initial background toroidal field
strengths?  As we will see, this last question has direct relevance to 
our stratified simulations, in which the net toroidal field within a given
region changes in time via shear and buoyancy.

Table~\ref{tbl:unstrat_sims} is a list of the unstratified shearing
box simulations.  The column labeled ``Resolution'' lists $N_x$, the number of
zones in one $H$.  The column labeled $\alpha$ gives the
averaged stress normalized by the averaged gas pressure,
\begin{equation}
\label{define_alpha}
\alpha \equiv \frac{\langle\langle\rho v_x\delta v_y - B_x B_y \rangle\rangle}{\langle\langle P \rangle\rangle},
\end{equation}

\noindent where the double brackets denote a time and volume average,
the volume average is calculated over the entire simulation domain, and
the time average is calculated from orbit 20 to the end of the run. Since
the gas is isothermal, $\langle P \rangle = \langle\rho\rangle\cs^2$.

\subsection{Resolving Physical Dissipation}

\begin{figure}
\includegraphics*[scale=0.7,angle=0]{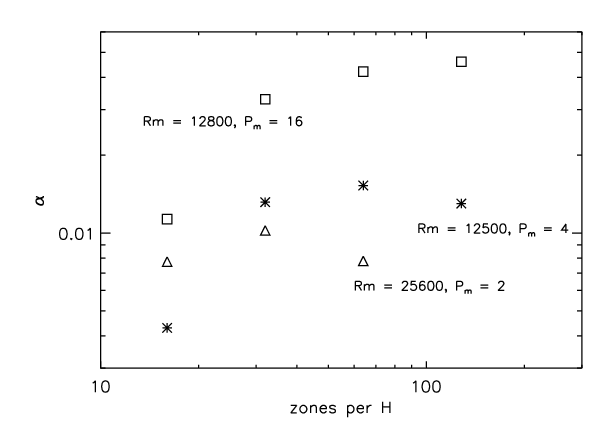}
\caption{Time- and volume-averaged stress parameter $\alpha$ as a
function of grid zones per $H$ in the unstratified SZ simulations; 
$\alpha \equiv
\langle\langle \rho v_x\delta v_y - B_xB_y \rangle\rangle/\langle\langle
P\rangle\rangle$, where the average is calculated over the entire
simulation domain and from 20 orbits to the end of the simulation.
Only simulations with sustained turbulence are plotted.  The squares
are runs with $Rm = 12800$, $\pr = 16$; asterisks are $Rm = 12500$, $\pr = 4$; and
triangles are $Rm = 25600$, $\pr = 2$.  By 32 zones
per $H$, the $\alpha$ values appear to be relatively close to the higher
resolution values.  }
\label{unstrat_convergence}
\end{figure}

We begin with a resolution study of unstratified simulations
with physical dissipation.  Figure~\ref{unstrat_convergence} shows
$\alpha$ as a function of resolution for three different $\pr$ values in
the SZ simulations.  The $N_x = 128$ data listed in the table
and plotted in the figure were taken from \cite{simon09b}.  Note that
those runs were similar to, but not the same as those done for this paper.
Here, the grid zone size is equal in all directions, $\Delta x = \Delta y = \Delta z$, whereas the
\cite{simon09b} runs had $\Delta x  = \Delta z = 0.4 \Delta y$.  Furthermore,
the calculations done in \cite{simon09b} were performed with the Roe
method for the Riemann solver, in contrast to the HLLD solver used here.

For the SZ simulations that have sustained turbulence, $\alpha$
appears to be converging with resolution.  More specifically, by 32 grid
zones per $H$, $\alpha$ is within a factor of $\sim 1.4$ of
the corresponding value at 128 zones per $H$.  In contrast, $\alpha$
in the $\pr = 16$ and $\pr = 4$ runs increases by about a factor of
3 going from 16 to 32 zones per $H$.  The $\pr = 2$ case shows a much
smaller change and the turbulence dies out for all resolutions with $Re =
6250$ and $\pr = 1$ in agreement with the higher resolution simulations
of \cite{fromang07b}.

\begin{figure}
\includegraphics*[scale=0.7,angle=0]{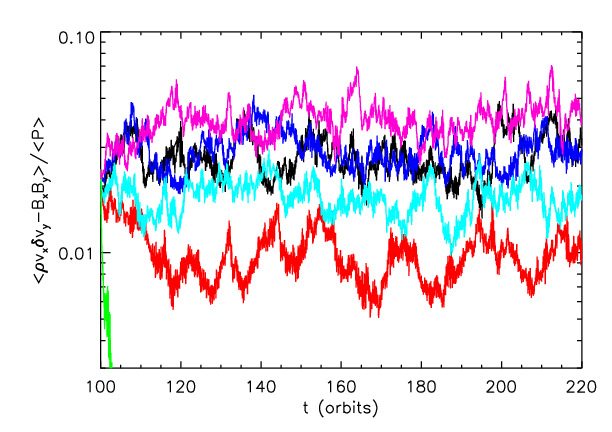}
\caption{Volume-averaged total stress normalized by the
the volume-averaged gas pressure as a function of time in orbits in the
unstratified, toroidal field FT simulations.  The black line corresponds
to $Rm = 3200$ and $\pr = 4$, magenta is $Rm = 6400$ and $\pr = 8$, green
is $Rm = 800$ and $\pr = 0.5$, light blue is $Rm = 3200$ and $\pr = 2$,
dark blue is $Rm = 6400$ and $\pr = 4$, and red is $Rm = 3200$ and $\pr
= 0.5$.  There is a clear increase in stress with increasing $\pr$.
For sufficiently low $Rm$, the turbulence decays.  }
\label{ft_unstrat_hist}
\end{figure}

\begin{figure}
\includegraphics*[scale=0.7,angle=0]{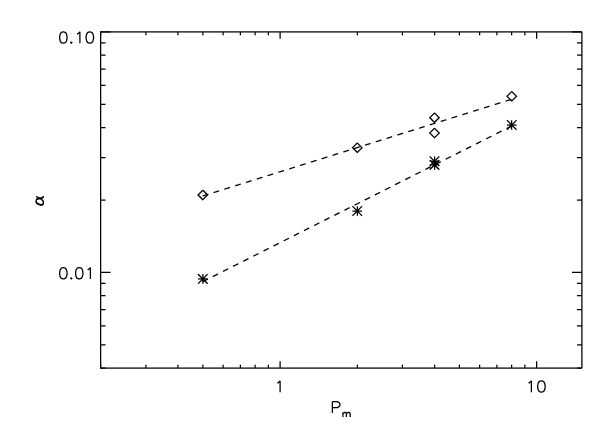}
\caption{ Time- and volume-averaged stress parameter $\alpha$
as a function of $\pr$ in the unstratified FT simulations
(asterisks) and the higher resolution, net toroidal field simulations
of \cite{simon09b} (diamonds). In the FT simulations, $\alpha \equiv
\langle\langle \rho v_x\delta v_y - B_xB_y \rangle\rangle/\langle\langle
P\rangle\rangle$, whereas in the higher resolution simulations, $\alpha
\equiv \langle\langle \rho v_x\delta v_y - B_xB_y \rangle\rangle/P_o$;
see \cite{simon09b}.  These definitions are roughly equivalent since
$\langle P \rangle \approx P_o$. For the FT simulations, the average is
calculated over the entire simulation domain and from 120 orbits to the
end of the simulation.  Only simulations with sustained turbulence are
plotted.  The dashed lines are linear fits to the data in log-log space.
Both resolutions show a clear $\pr$ dependence, but this dependence is
steeper at the lower resolution.  }
\label{ft_res}
\end{figure}

Next, we explore the influence of physical dissipation terms on the
FT initial field configuration, using 32 zones per $H$.  The time
history of the volume-averaged stress for these runs is displayed
in Fig.~\ref{ft_unstrat_hist}.  There is a clear dependence on the
dissipation parameters and on $\pr$ in particular.  For large enough
resistivity (i.e., low $Rm$), the turbulence decays; the critical $Rm$
value is $\sim 1000$, in agreement with the higher resolution simulations
of \cite{simon09b}.  In Fig.~\ref{ft_res}, we plot the time-averaged
$\alpha$ values, averaged from orbit 120 to the end of the simulation,
versus $\pr$ for these runs (asterisks).  Note that 32FTRm3200Pm4
and 32FTRm6400Pm4 have different $Rm$ values, but the same $\pr$ and
nearly the same saturation level.  We also plot $\alpha$ from the higher
resolution simulations of \cite{simon09b} (see their Table 3).  The dashed
lines are linear fits to the data in log-log space.  Assuming $\delta$
in $\alpha \propto \pr^\delta$ we find $\delta=0.54$ for 32 grid zones
per $H$, and $\delta = 0.33$ for 128 grid zones per $H$; the $\pr$
dependence is steeper at the lower resolution.  Furthermore, all of the
$\alpha$ values for the higher resolution simulations are larger than
the corresponding lower resolution simulations.

Both the zero net flux and net toroidal flux results suggest that moderate
resolutions, i.e., 32 grid zones per $H$, may be sufficient to capture
the general effects of changing $\nu$ and $\eta$, at least for the range
of $Re$, $Rm$, and $\pr$ values considered here.  This is not to say that
everything is converged at this resolution.  Indeed, Fig.~\ref{ft_res}
shows a noticeable resolution effect.  Full convergence likely requires
a sufficiently high resolution to ensure that the effective numerical
dissipation scale is below the viscous and resistive dissipation scales
\cite[see e.g.,][]{fromang07b,simon09a,simon09b}, but the general
dependence of $\alpha$ on dissipation parameters appears to be captured
even at 32 zones per $H$.  This conclusion is also supported by the recent
results of \cite{flaig10}, which show that $\sim 30$ zones per $H$ in the
vertical direction may be sufficient to characterize the turbulence in
their stratified shearing box simulations.  This is an important point,
as available computational resources limit us to this resolution
in carrying out the comprehensive study of physical dissipation effects presented
in this work.

\subsection{Dissipation and Initial Field Strength}

Because the fastest growing MRI wavelength is proportional to $\va$,
the background field strength can play a significant role in the
outcome of a shearing box simulation.  For example, 
\cite{longaretti10} found that the dependence of angular momentum
transport on $Rm$ and $\pr$ becomes steeper for weaker background
(vertical) fields in unstratified shearing boxes.

\begin{figure}
\includegraphics*[scale=0.7,angle=0]{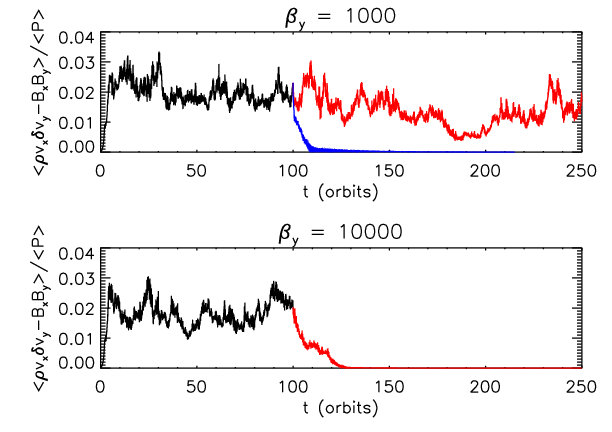}
\caption{
Volume-averaged total stress normalized by the
volume-averaged gas pressure as a function of time in orbits for a
series of unstratified shearing box simulations. The volume-average is
done over the entire simulation domain.  In each plot, the black line
is from a simulation with only numerical dissipation.  The colored lines
are simulations with physical dissipation, initiated from the numerical
dissipation run at orbit 100.  The red lines correspond to $Rm = 3200$,
$\pr = 2$ and the blue line is $Rm = 1600$, $\pr = 1$. The top panel is
initiated with a background toroidal field characterized by $\beta =
1000$, and the bottom panel has $\beta = 10000$.  The weaker toroidal
field appears to be killed off at a lower resistivity compared to the
stronger background field.
}
\label{beta_hist}
\end{figure}

Here we consider the effects of different toroidal field strengths on
the FT initial condition in the unstratified shearing box.  We consider
$\beta_y = 1000$ and $\beta_y = 10000$ runs both with and without physical
dissipation (see Table~\ref{tbl:unstrat_sims} and \S~\ref{method}). The
time history of the volume-averaged stresses for these simulations is
shown in Fig.~\ref{beta_hist}.  For $\beta_y = 1000$, the turbulence
survives at $Rm = 3200$ but dies at $Rm = 1600$, whereas for $\beta_y =
10000$, the turbulence dies at $Rm = 3200$.  It would seem that for
each increase in $\beta_y$ by a factor of 10, the critical $Rm$ value
increases by roughly a factor of 2; it becomes easier to kill off the
MRI with a lower resistivity as the background toroidal field is weakened.

The time- and volume-averaged Elsasser numbers, calculated via
equation~(\ref{elsasser}), are 2.3 and 137 for the $\beta =1000$, $Rm =
1600$ and 3200 models respectively. The time average is from orbit 140 to
the end of the calculation, and since the turbulence decays, the average
$\Lambda$ for $Rm = 1600$ agrees with that given by the initial toroidal
field value.  For the $\beta=10000$, $Rm = 3200$ model, the Elsasser
number is 0.45, again, calculated from either the initial toroidal
field strength or from equation~(\ref{elsasser}) after the turbulence has
completely decayed. In both cases the initial poloidal field has $\beta_p
= 1600$, corresponding to an initial poloidal $\Lambda$ on order unity for
these $Rm$ values.  These results are consistent with the $\Lambda$ values
calculated in \cite{simon09b}; $\Lambda \lesssim 100$ leads to decay.

The time- and volume-averaged $Q_y$ values are $Q_y = 34$ and $Q_y = 33$
for 32FTNum$\beta$1000 and 32FTNum$\beta$10000, respectively.  The time
average is done from orbit 20 to 110. The toroidal field MRI appears to
be quite well-resolved in both simulations.  Averaging over the same time
interval, we find that $Q_z = 7.3$ for 32FTNum$\beta$1000 and $Q_z = 6.9$
for 32FTNum$\beta$10000; the vertical field MRI is marginally resolved.
Note that these $Q$ values are calculated via the saturated state of
these runs.  Indeed, the initial $Q$ values are substantially lower and
either marginally or under-resolved.  It is not clear which $Q$ values
are a better representation of how well-resolved the MRI actually is; the
initial values correspond to the background field which ultimately drives
the MRI, but other modes may become important in driving the MRI in the
fully nonlinear state.  In any case, despite the low initial $Q$ values,
sustained turbulence develops when explicit dissipation terms are not
included, suggesting that at least some MRI modes are resolved. It is only
when resistivity is turned on that decaying turbulence is observed. The
initial low $Q$ values may introduce uncertainty for the specific values
of the critical resistivity, but the relationship between critical
resistivity and initial field strength is likely to hold nevertheless.

% SECTION 4

\section{Vertically Stratified Simulations}
\label{strat_sims}

\subsection{Baseline Simulations}

We now turn to a series of simulations that investigate the effects of
physical dissipation in vertically stratified, isothermal disks.  The
various combinations of runs are summarized in Table~\ref{tbl:strat_sims}.
Some time- and volume-averaged values of various quantities are given
in Table~\ref{tbl:sat_char}.   The baseline calculations without physical dissipation
are done at two resolutions: 32Num which uses 32 grid zones per $H$,
and 64Num which uses 64. 32Num is run to a total integration time of
1058 orbits. It has sustained turbulence at a level of $\alpha = 0.028$,
calculated via equation~(\ref{define_alpha}), time-averaging from orbit
20 until the end of the calculation, and volume-averaging over all $x$
and $y$ and for $|z| \le 2 H$.  In Model 64Num, $\alpha = 0.022$, where
the time average is from orbit 20 until the end of the simulation at
159 orbits.

\begin{figure}
\begin{center}
\subfigure{
\includegraphics[width=6.25in,height=3.125in,angle=0]{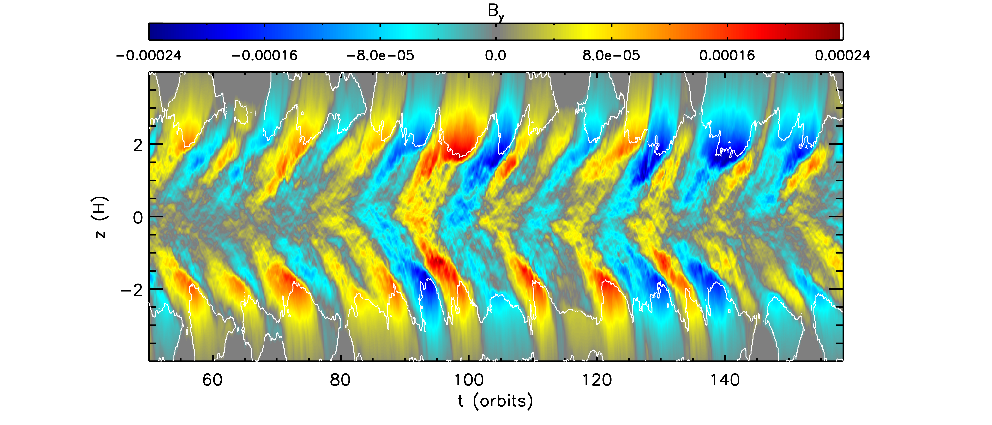}
}
\subfigure{
\includegraphics[width=6.25in,height=3.125in,angle=0]{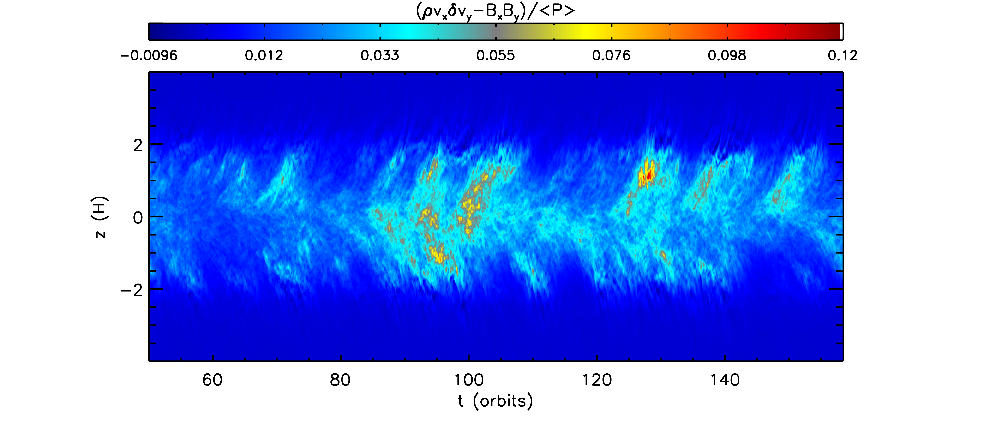}
}
\end{center}
\caption{  
Space-time diagram of the horizontally averaged $B_y$ (top panel) and
total stress normalized by the volume-averaged gas pressure (bottom
panel).  The volume-average is done for all $x$ and $y$ and for $|z|
\le 2 H$.  The white contours on the top panel denote where $\beta$
goes from greater to less than unity.  The horizontally
averaged $B_y$ appears to rise vertically into the upper $z$ layers,
being replaced in the mid-plane region by $B_y$ of the opposite sign.
The rise speed of the field increases after $|z| \sim 2 H$ is reached. The
sign flipping in $B_y$ has a period of $\sim 10$ orbits.
}
\label{64i_sttz}
\end{figure}

One particularly useful diagnostic is the space-time diagram of
horizontally averaged quantities.  Figure~\ref{64i_sttz} shows space-time
plots of horizontally averaged $B_y$ and pressure-normalized total
stress for a 100 orbit period in the 64Num simulation.  From this,
we see that $B_y$ changes sign periodically and rises above the
equatorial plane. The behavior of $B_y$ is roughly mirror symmetric around the equator and
has a period of $\sim 10$ orbits at both resolutions.
This has been seen in previous vertically stratified shearing boxes
\cite[e.g.,][]{brandenburg95,stone96a,hirose06,guan10,shi10,gressel10,davis10},
with a variety of initial fields, numerical resolutions and codes,
and it has even been observed in global simulations
\cite[e.g.,][]{fromang06b,dzyurkevich10,oneill10}.  It is apparently a generic property of MRI-driven
turbulence in the presence of vertical gravity.

The white contours in the top panel of Fig.~\ref{64i_sttz} indicate
where the gas $\beta$ value switches from greater than to less than unity.
For $|z| \gtrsim 2-2.5 H$, $\beta < 1$, except for some regions very near
the vertical boundaries where $\beta > 1$ as a result of the absence of
magnetic field.  The region around $|z| = 2 H$ is also where the fluid
becomes completely buoyantly unstable.  Using the criterion of \cite{newcomb61}
as outlined in \cite{guan10}, the gas is buoyantly stable if

\begin{equation}
\label{buoyancy_stability}
\left|\frac{d\rho}{dz}\right| > \left|\frac{\rho^2 g}{\gamma P}\right|
\end{equation}

\noindent where $\gamma = 1$ because the gas is isothermal.  According to
this criterion applied to the simulation data,
the fluid is buoyantly unstable for $|z| \gtrsim 2 H$. 
Consistent with this, the space-time plot shows that the
field rises faster for $|z| \gtrsim 2 H$. For $|z| \lesssim 2 H$,
there are regions of instability as well as stability, and it is within
this region that the MRI is active as indicated by the presence of
significant stress.  Indeed, the total stress drops off rapidly near $|z| \sim 2H$.
 It appears that the marginal buoyancy stability coupled with the
MRI-induced turbulence leads to a slower rise of magnetic field until
$|z| \sim 1.5$--$2 H$, where the gas then becomes completely buoyantly unstable.
These results are consistent with the recent ZEUS calculations of
\cite{guan10} with large radial extent as well as with \cite{shi10}
using a version of ZEUS that includes radiation physics and total energy
conservation.  Compare, e.g., the top panel of Fig.~\ref{64i_sttz}
to Fig.~6 in \cite{shi10}.  

\begin{figure}[p]
\begin{minipage}[!ht]{8cm}
\begin{center}
\includegraphics[width=1\textwidth,angle=0]{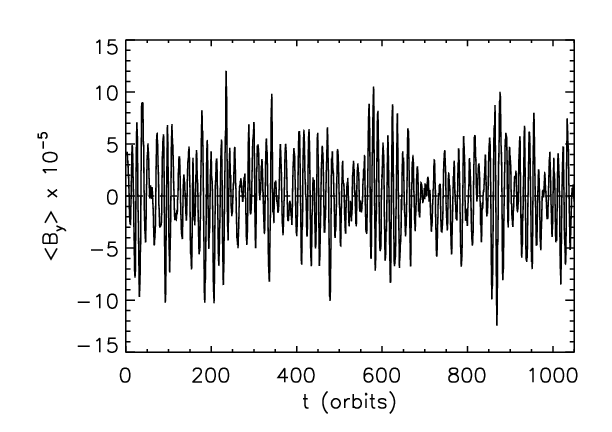}
\end{center}
\end{minipage}
\begin{minipage}[!ht]{8cm}
\begin{center}
\includegraphics[width=1\textwidth,angle=0]{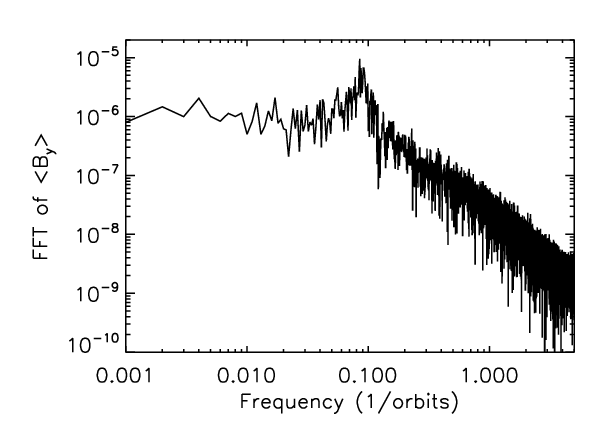}
\end{center}
\end{minipage}
\caption{  
Left: Time evolution of volume-averaged $B_y$ in 32Num. 
The volume average is done for all $x$ and $y$ and for $|z| \le 0.5 H$.  The dashed line corresponds to $\langle B_{x,y}\rangle = 0$.
Right: Temporal power spectrum of $\langle B_y\rangle$ from the left plot, calculated from orbit 50 to 1050. 
The 10 orbit period oscillations in $\langle B_y\rangle$ are immediately apparent in both plots, particularly as the peak in the power
spectrum.  The 10 orbit 
oscillations are modulated on longer timescales, ranging from tens to hundreds of orbits.
}
\label{ideal_by}
\end{figure}

\begin{figure}
\includegraphics[scale=0.7,angle=0]{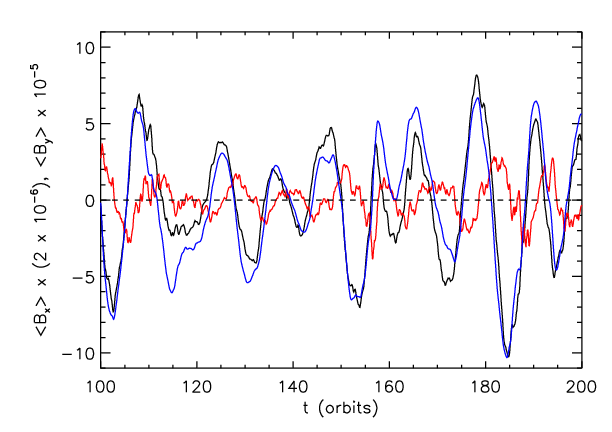}
\caption{  
Time evolution of volume-averaged field components for part of 32Num.  Red is $\langle B_x\rangle$, black is $\langle B_y\rangle$, and blue is $\langle B_y\rangle$ as calculated from
$\langle B_x\rangle$ using a simple $\alpha$--$\Omega$ dynamo model discussed in the text.  The volume average is done for all $x$ and $y$ and for $|z| \le 0.5 H$.  The dashed line corresponds to $\langle B_{x,y}\rangle = 0$.
$\langle B_x\rangle$ has been multiplied by a factor of 5 relative to $\langle B_y\rangle$ to make a more direct comparison possible.
The variations in $\langle B_x\rangle$ are accompanied by variations in $\langle B_y\rangle$, which are offset in time, and the dynamo model shows that the evolution of $\langle B_y\rangle$
is controlled by shear of radial field and buoyant removal of toroidal field.
}
\label{ideal_bx_by}
\end{figure}

Figure~\ref{ideal_by} shows the volume-averaged toroidal field within $|z|
\le 0.5 H$ as a function of time. The RMS field fluctuations averaged
within this region are roughly a factor of $\sim 4$ times larger; the
turbulent fluctuations dominate over the mean, but not significantly
so. The temporal oscillations in the mean field are apparent.  The right
figure shows the temporal power spectrum, revealing the dominant $\sim
10$ orbit period.  The oscillation amplitude appears to be modulated on
longer timescales, ranging from tens to hundreds of orbits.  Furthermore,
the averaged radial field appears to exhibit the same 10 orbit cyclic
behavior as the toroidal field, but with a slight temporal lag, as
shown in Fig.~\ref{ideal_bx_by}.  The behavior of the mean radial and
toroidal fields resembles the $\alpha$--$\Omega$ dynamo model derived in
\cite{guan10}.  Specifically, they write simplified evolution equations
for volume-averaged field components,

\begin{equation}
\label{by_alpha_dyn}
\frac{d \langle B_y\rangle}{dt} = -q \Omega \langle B_x\rangle - \frac{|\va|}{2H}\langle B_y\rangle + \frac{\alpha_1}{2 H} \langle B_x\rangle,
\end{equation}

\begin{equation}
\label{bx_alpha_dyn}
\frac{d \langle B_x\rangle}{dt} = - \frac{|\va|}{2H}\langle B_x\rangle
- \frac{\alpha_2}{2 H} \langle B_y\rangle .
\end{equation}

\noindent The first term on the right hand side of
equation~(\ref{by_alpha_dyn}) is the background shear acting on the radial
field.  The second term is the loss of toroidal field due to buoyant rise,
estimated to have a characteristic buoyant velocity equal to the toroidal
field $\alf$ speed.  The third term is the $\alpha$-dynamo term coupling
$B_x$ to the evolution of $B_y$.  Equation~(\ref{bx_alpha_dyn}) is similar
except there is no shear term and the toroidal and radial field components
have been flipped with respect to equation~(\ref{by_alpha_dyn}).  Also,
in general, $\alpha_1 \ne \alpha_2$.  Note that while we follow
the notation and general approach of \cite{guan10} here, \cite{gressel10}
has also constructed an $\alpha$--$\Omega$ model, which incorporates a
helicity-based dynamo mechanism and reproduces the observed behavior of
the horizontally averaged field components.

\cite{guan10} numerically integrated this set of equations and found
a solution that looks strikingly similar to the red and black curves in
Fig.~\ref{ideal_bx_by}.  The value of $\alpha_{1,2}$ sets the oscillation
frequency in the mean field; $\alpha_1 = \alpha_2 = -0.01\Omega H$
reproduces the 10 orbit variability observed in stratified shearing boxes.
Here we numerically integrate equation (\ref{by_alpha_dyn}) using the
simulation data for $\langle B_x\rangle$ (the red curve) and the initial
condition for $\langle B_y\rangle$ taken from $\langle B_y\rangle$ at
$t = 0$.  We have set $\alpha_1 = 0$ so that the $\langle B_y \rangle$
evolution is controlled entirely by shear and buoyancy.  The result is
the blue curve in the figure.  The agreement between the actual evolution
of $\langle B_y\rangle$ and the $\alpha$--$\Omega$ dynamo model suggests
that the evolution of the mean toroidal field within the mid-plane region
is almost completely controlled by the shearing of radial field and the
buoyant removal of the generated toroidal field.

The remaining question then is, what creates the mean radial field?
In the dynamo model, the $\alpha_2$ term is essential, but
what mechanism is responsible for $\alpha_2 \ne 0$?  The most likely
candidate is MRI turbulence; turbulent fluctuations create EMFs that
generate poloidal field \cite[e.g.,][]{brandenburg95,davis10,gressel10},
but the details of this are still not well understood.

\begin{figure}
\includegraphics*[scale=0.7,angle=0]{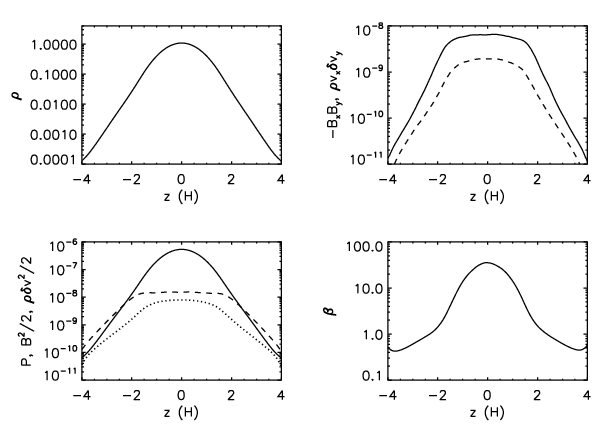}
\caption{  
Time- and horizontally-averaged vertical distributions for various quantities in 32Num.  Upper left: gas density; upper
right: Maxwell (solid) and Reynolds (dashed) stresses; lower left: gas pressure (solid), magnetic energy (dashed),
and kinetic energy (dotted); lower right: gas $\beta$ defined as the time- and horizontally-averaged gas pressure
divided by the time- and horizontally-averaged magnetic energy density.
The time average is done from orbit 100 to the end of the simulation.
The stress and magnetic energy are relatively flat for $|z| \lesssim 1.5 H$ but drop off substantially for larger $|z|$. 
Outside of $|z| \sim 2 H$, the magnetic energy dominates over gas pressure.
}
\label{z_profile_ideal}
\end{figure}

In Fig.~\ref{z_profile_ideal}, we plot the time- and horizontally-averaged
vertical distributions for various quantities in 32Num. The time
average runs from orbit 100 to the end of the calculation. The figure
shows that the stress drops off rapidly near $|z| \sim 1.5 H$. The
shape of the distribution is generally the same for both Maxwell and
Reynolds stresses, with the Maxwell stress always greater than the
Reynolds stress by a factor that varies from 2.3 to 5.6 depending
on $z$; this factor is $\sim 4$ when averaged over all $z$, in
agreement with unstratified simulations \cite[e.g.,][]{hawley95a}.
The gas pressure maintains an approximately Gaussian distribution
consistent with hydrostatic equilibrium, while the magnetic
pressure is subthermal and relatively flat for all $|z| \lesssim 2 H$.  
The magnetic field becomes superthermal
for $|z| > 2 H$, although its magnitude continues to decrease with height.
These results are consistent with previous studies of isothermal disks
\cite[e.g.,][]{stone96a,miller00,guan10}.  Interestingly, the vertical
structure of the turbulence is also consistent with local simulations containing
radiation physics \cite[]{hirose06,krolik07a,hirose09a,flaig10} and
global simulations \cite[]{hawley01b,fromang06b}, though
we do not observe the double peak profile in the stress as seen in the
simulations of \cite{hirose09a} and \cite{flaig10}.

\begin{figure}
\includegraphics*[scale=0.6,angle=270]{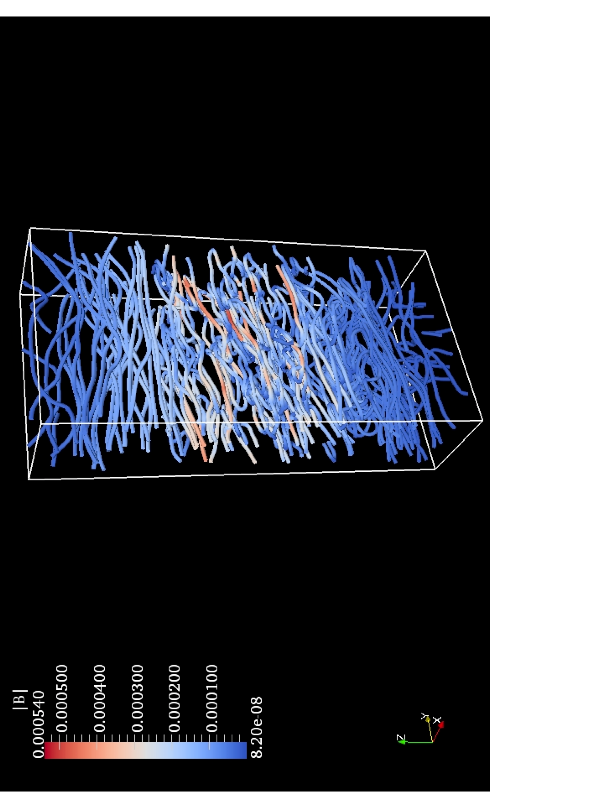}
\caption{  
Magnetic field structure at $t = 100$ orbits in 32Num, produced
via a stream line integration.  The field strength (in code units) is displayed
via color and not the density of the field lines.  The magnetic field
has a primarily toroidal structure but has a smaller, tangled structure in the $x$ and $z$
directions.
}
\label{ideal_field_lines}
\end{figure}

Figure~\ref{ideal_field_lines} examines the three-dimensional structure of
the magnetic field in the fully turbulent gas using streamline integration
for 32Num at $t = 100$ orbits.  The field strength is denoted by color
rather than field line density.  Within $|z| \lesssim 2 H$, the field is
primarily toroidal but has a smaller scale, tangled structure in the $x$
and $z$ directions.  Very near the vertical boundaries, however, the
field appears to develop larger poloidal components.
Utilizing snapshots taken throughout the evolution of 32Num, we find that this
is typical of the saturated state, except for at $t = 550$ orbits, in
which the field near the boundaries is primarily vertical.   Figure 16
of \cite{hirose06} shows the field in a shearing box with a vertical
domain twice as large as 32Num.  In their simulation, the field at
$|z| = 4 H$ is mainly toroidal while the field structure at $|z| = 4 H$
in 32Num resembles the field structure near $|z| = 8 H$ in their
simulation.  This suggests that the vertical outflow boundary conditions
are influencing the field structure very near the boundaries.  Away from
the vertical boundaries, however, the magnetic field in 32Num appears to have a
very similar structure to that in \cite{hirose06}.

\begin{deluxetable}{l|cccccccc}
\rotate
\tabletypesize{\scriptsize}
\tablewidth{0pc}
\tablecaption{Saturation Characteristics for Sustained Turbulence Runs\label{tbl:sat_char}}
\tablehead{
\colhead{Quantity}&
\colhead{32Num\tablenotemark{a}}&
\colhead{32Rm3200Pm4\tablenotemark{a}}&
\colhead{32Rm6250Pm1\tablenotemark{a}}&
\colhead{32Rm6400Pm4\tablenotemark{a}}&
\colhead{32Rm6400Pm8\tablenotemark{a}}&
\colhead{64Num\tablenotemark{b}}&
\colhead{64Rm3200Pm2\tablenotemark{a}}&
\colhead{64Rm6400Pm4\tablenotemark{a}}    } 
\startdata
$\maxwell/\pavg$ & 0.022 & 0.023 & 0.018 & 0.025 & 0.033 & 0.018 & 0.027 & 0.025 \\
$\reynolds/\pavg$ & 0.0058 & 0.0051& 0.0046 & 0.0061 & 0.0071 & 0.0044 & 0.0060 & 0.0055 \\
$\magnetic/\pavg$ & 0.059 & 0.069 & 0.050 & 0.069 & 0.089 & 0.045 & 0.074 & 0.063 \\
$\magx/\pavg$ & 0.0066 & 0.0059 & 0.0051 & 0.0069 & 0.0084 & 0.0061& 0.0079 &  0.0072 \\
$\magy/\pavg$ & 0.049 & 0.060 & 0.043 & 0.058 & 0.077 & 0.036 & 0.062 & 0.052 \\
$\magz/\pavg$ & 0.0031 & 0.0028 & 0.0024 & 0.0032 & 0.0039 & 0.0030 & 0.0040 & 0.0035 \\
$\pertkin/\pavg$ & 0.024 & 0.021 & 0.020 & 0.024 & 0.027 & 0.020 & 0.025 & 0.022 \\
$\kinx/\pavg$ & 0.010 & 0.0088 & 0.0085 & 0.010 & 0.011 & 0.0079 & 0.010 & 0.0094 \\
$\pertkiny/\pavg$ & 0.0079 & 0.0066 & 0.0062 & 0.0078 & 0.0089 & 0.0070 & 0.0084 & 0.0075 \\
$\kinz/\pavg$ & 0.0059 & 0.0057 & 0.0052 & 0.0061 & 0.0068 & 0.0047 & 0.0065 & 0.0055 \\
$\maxwell/\reynolds$ & 3.77 & 4.55 & 3.87 & 4.18 & 4.58 & 4.03 & 4.47 & 4.60 \\
$\maxwell/\magnetic$ & 0.37 & 0.34 & 0.36 & 0.37 & 0.37 & 0.40 & 0.36 & 0.40 \\
\enddata
\tablenotetext{a}{Volume averaged for $|z| \le 2 H$ and time averaged from orbit 150 to the end of the simulation.}
\tablenotetext{b}{Volume averaged for $|z| \le 2 H$ and time averaged from orbit 50 to the end of the simulation.}
\end{deluxetable}

\subsection{High and Low Turbulence States}

Having established the baseline simulations without physical dissipation,
we now turn to the main focus of our present work:
the effect of $\nu$ and $\eta$ on vertically stratified MRI turbulence.

\begin{figure}[p]
\begin{minipage}[!ht]{8cm}
\begin{center}
\includegraphics[width=1\textwidth,angle=0]{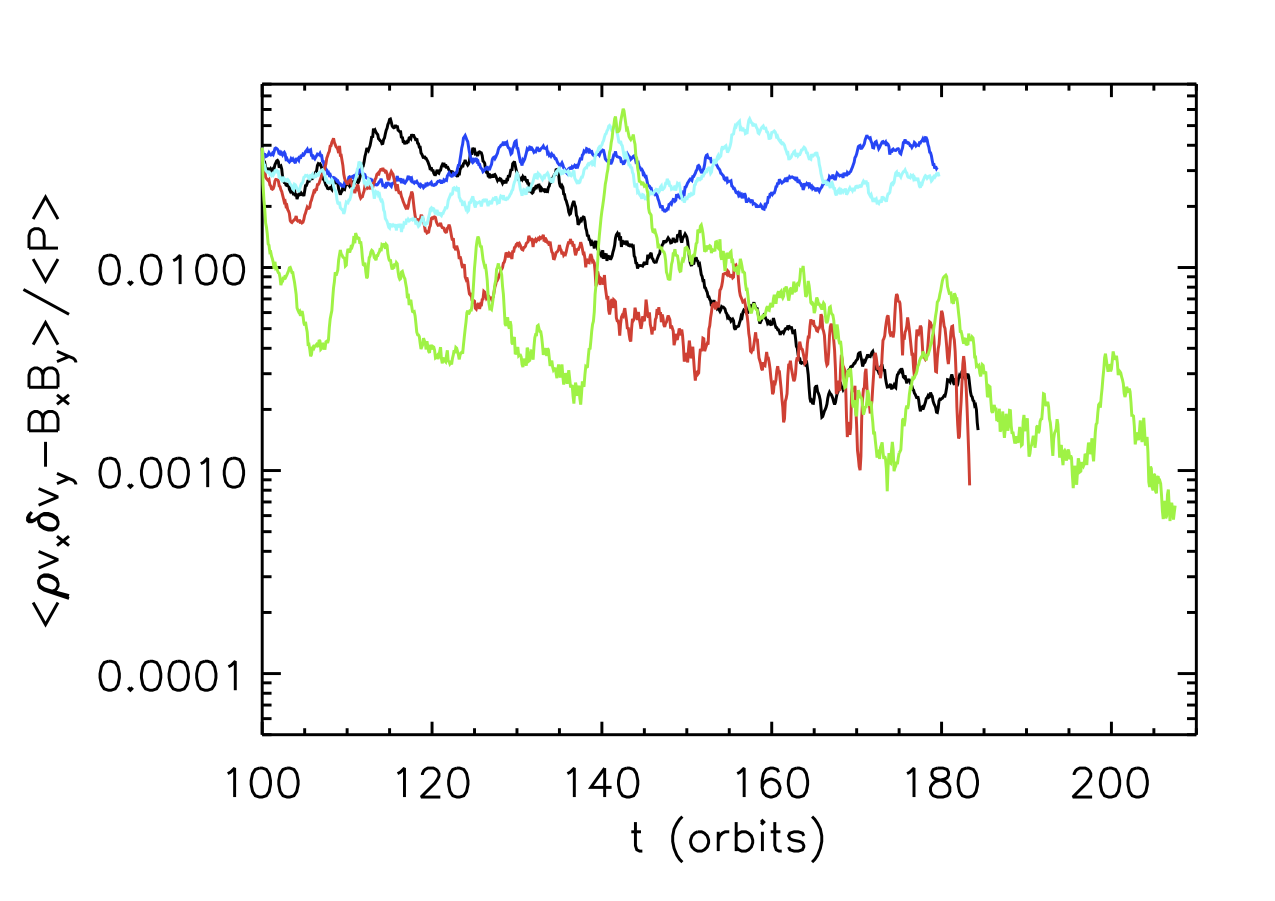}
\end{center}
\end{minipage}
\begin{minipage}[!ht]{8cm}
\begin{center}
\includegraphics[width=1\textwidth,angle=0]{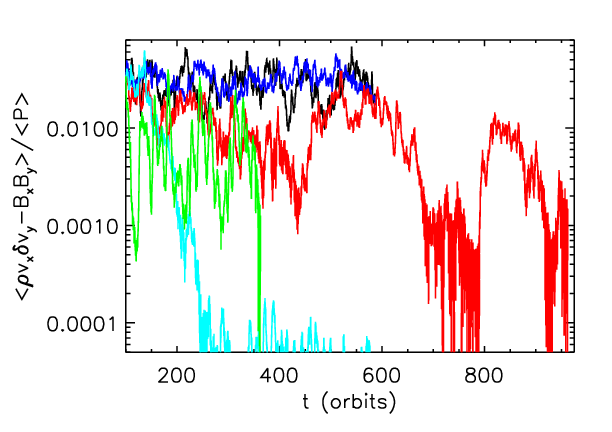}
\end{center}
\end{minipage}
\caption{Volume-averaged total stress normalized by the volume-averaged gas pressure
as a function of time in orbits in the vertically stratified simulations at 64 (left) and 32 (right) grid zones per $H$. The 
volume-average is done for all $x$ and $y$ and for $|z| \le 2 H$. The black line corresponds to 
$Rm = 3200$ and $\pr = 4$, green is $Rm = 800$ and $\pr = 0.5$, 
light blue is $Rm = 3200$ and $\pr = 2$, dark blue is $Rm = 6400$ and $\pr = 4$, and red is $Rm = 3200$ and $\pr = 0.5$.
Some of the simulations appear to undergo periods of low stress followed by higher stress, occurring
on very long timescales of $\sim 100$ orbits in some cases.
}
\label{strat_hist}
\end{figure}

The Maxwell and Reynolds stress time evolution for the simulations
performed with 64 zones per $H$ is shown in the left panel of
Fig.~\ref{strat_hist}.  Runs 64Rm3200Pm0.5, 64Rm800Pm0.5, and 64Rm3200Pm4
have decreasing turbulence levels, while turbulence is sustained at a
statistically constant value in the remaining simulations.  Furthermore,
64Rm800Pm0.5 undergoes alternating periods of low and high stress,
though the overall trend is downward with time.

The right panel of Fig.~\ref{strat_hist} is the stress evolution for the
equivalent simulations with 32 zones per $H$, shown for a much longer
time period than the 64 zone runs.  There is considerable variability
on long timescales. 32Rm3200Pm0.5 in particular exhibits alternating
periods of low and high stress levels, occurring on $\sim 100-200$
orbit timescales.  The more viscous and resistive run, 32Rm800Pm0.5,
shows similar variability but on a much shorter timescale of $\gtrsim
10$ orbits.  Turbulence in 32Rm3200Pm2 appears to have leveled off at
a smaller value without any indication of regrowth to the higher state.
Both 32Rm6400Pm4 and 32Rm3200Pm4 remain at relatively high stress
levels, which are very similar between the two runs, likely a result of
having the same $\pr$.

\begin{figure}[p]
\begin{minipage}[!ht]{8cm}
\begin{center}
\includegraphics[width=1\textwidth,angle=0]{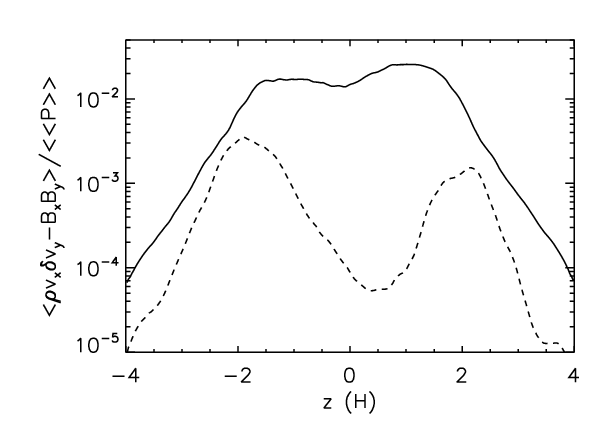}
\end{center}
\end{minipage}
\begin{minipage}[!ht]{8cm}
\begin{center}
\includegraphics[width=1\textwidth,angle=0]{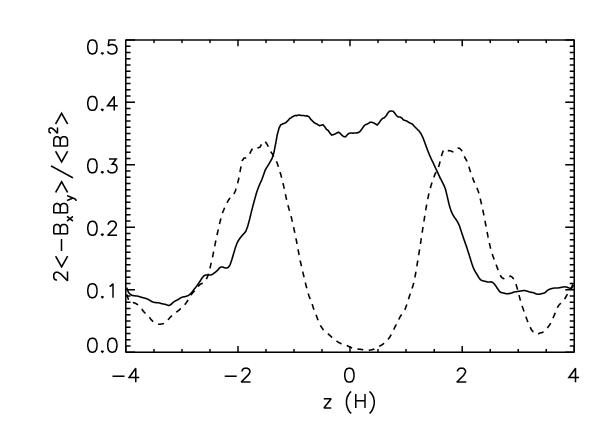}
\end{center}
\end{minipage}
\caption{  
Left: Time- and horizontally-averaged total stress as a function of $z$
for 32Rm3200Pm0.5. The stress is normalized by the time- and volume-averaged gas pressure, where the volume average is done for
all $x$ and $y$ and for $|z| \le 2 H$.  Right: Ratio of  time- and horizontally-averaged Maxwell stress
to time- and horizontally-averaged magnetic energy in 32Rm3200Pm0.5. In both plots, the solid line corresponds to a time average from 500 to 570 orbits, which is a state of high turbulence, and 
the dashed line is a time average from 700 to 770 orbits, a low turbulence state. During the high state,
the stress is relatively flat for $|z| \le 2 H$. In the low state, the stress is smaller within the mid-plane region and peaks near $|z| \sim 2 H$.  
}
\label{high_low_compare}
\end{figure}

Turning to the high and low states in 32Rm3200Pm0.5, 
the left panel of Fig.~\ref{high_low_compare} shows the vertical profile
for the horizontally averaged total stress during a high turbulence
state (solid line, time averaged from 500 to 570 orbits) and a low
turbulence state (dashed line, time averaged from 700 to 770 orbits).
The high state has considerable stress within the
mid-plane region ($|z| \lesssim 2 H$), whereas in the low state, the
stress is peaked near $|z| \sim 2 H$.  Note that there is still a nonzero
stress in the mid-plane region even during the low state.   The Reynolds
stress is nonzero and relatively flat throughout the mid-plane, whereas the
Maxwell stress drops off dramatically at the mid-plane.  
What Maxwell stress is present in the low states seems to derive
from the presence of residual radial and toroidal field. (This effect was also noted by \cite{turner07} in the dead zone
regions of their shearing box calculations). One
characteristic of MRI turbulence is the ratio of the Maxwell stress
to the magnetic pressure, which is typically on order 0.4 \citep[][and
Table~\ref{tbl:sat_char}]{hawley95a, blackman08}. The vertical profile of
this ratio is shown in the right panel of Fig.~\ref{high_low_compare}.
This ratio is between 0.3 and 0.4 for $|z| \lesssim 1.5 H$ in the high
state and near $|z| \sim 2 H$ in the low state; in the mid-plane
region during the low state, the ratio is much smaller and approaches
zero.

These differences between the high and low states are seen in the other
simulations that exhibit this variability (except for 32Rm800Pm0.5,
in which the variability occurs on such a short timescale that temporal
averaging becomes difficult).  In summary, during the high state, the MRI
is fully active, producing turbulent stresses for $|z| \lesssim 2 H$. In
the low state, the turbulence in the mid-plane region has died down,
leaving weaker Reynolds stresses.  MRI-driven turbulence is present,
however, in narrow regions near $|z| \sim 2 H$.  This is similar to the
behavior seen in \cite{fleming03} where active layers above and below the
mid-plane drive Reynolds stresses in the equatorial dead zone.

Is the variability between high and low turbulence an artifact of using
a relatively low resolution in these calculations? Several pieces of
evidence suggest this is not the case.  First of all, both resolutions
for $Re = 1600$, $\pr = 0.5$ show the same variable stress behavior
on $\gtrsim 10$ orbit timescales.  Secondly, even at 32 zones per $H$,
dissipation coefficients play a significant role in determining the stress
level, as shown above. The fact that the low resolution, unstratified
simulations show constant saturation levels for sufficiently small $\eta$
whereas vertically stratified simulations exhibit this variability for
the same parameters suggests that the variability is a direct result of
adding in vertical gravity.  Thirdly, this variability was seen in the
simulations of \cite{davis10}, which were run at a higher resolution
of 64 zones per $H$.  We note, however, that while \cite{davis10} used
the same numerical algorithm as in this work, they used a different
initial magnetic field configuration and vertical boundary conditions.

\begin{figure}[p]
\begin{minipage}[!ht]{8cm}
\begin{center}
\includegraphics[width=1\textwidth,angle=0]{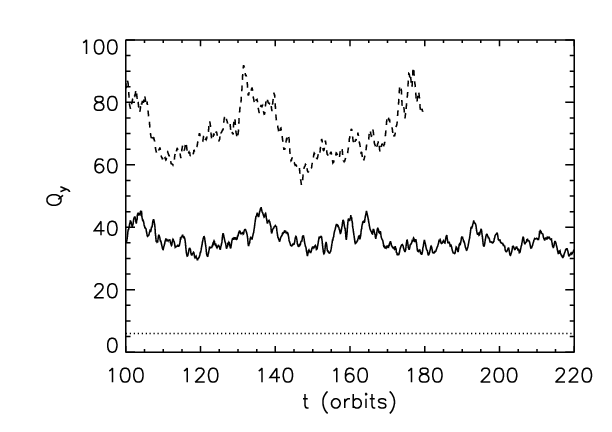}
\end{center}
\end{minipage}
\begin{minipage}[!ht]{8cm}
\begin{center}
\includegraphics[width=1\textwidth,angle=0]{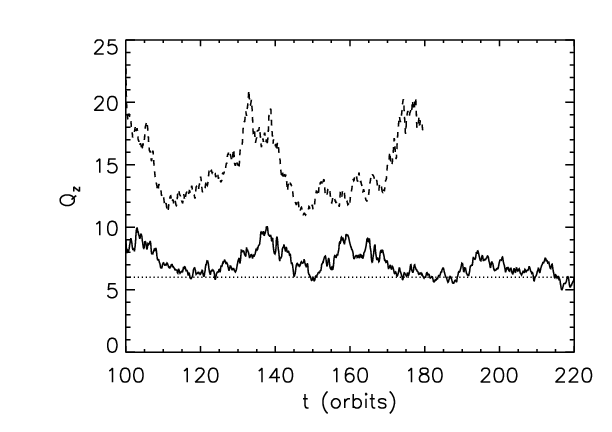}
\end{center}
\end{minipage}
\caption{
Quantitative measurement of how well resolved the MRI is in the toroidal ($Q_y$; left plot) and
vertical ($Q_z$; right plot) directions as a function of time.  The solid line corresponds to run 32Rm6400Pm4 and
the dashed line is 64Rm6400pm4.  The dotted horizontal line corresponds to $Q = 6$, below which the MRI
is considered to be under-resolved \cite[]{sano04}.
$Q_i$ is calculated using the volume average of the magnetic energy and gas density for all $x$ and $y$ and for $|z| \le 0.5 H $ (see text).
The toroidal field MRI at both resolutions as well as the vertical field MRI at the higher
resolution appear to be reasonably well-resolved. However, at the lower resolution, the vertical field MRI is only marginally resolved.
}
\label{q_re1600pm4}
\end{figure}

Finally, we examine the $Q$ values (\ref{q_resolve}) using an $\alf$
speed defined by (\ref{alfven_q}) and a volume average for all
$x$ and $y$ and for $|z| \le 0.5 H$.  Figure~\ref{q_re1600pm4} shows
$Q_y$ and $Q_z$ as a function of time for 32Rm6400Pm4 and 64Rm6400pm4.
The results suggest that the toroidal field MRI is quite well-resolved,
but that the vertical field MRI may be only marginally resolved for the lower
resolution simulation.  The higher resolution $Q$ values are roughly
a factor of 2 larger than the lower resolution $Q$ values, which is
simply a result of the change in the grid zone size;
the turbulent saturation level is roughly the same between the two
resolutions.  This result, coupled with the somewhat low $Q_z$ value,
suggests that the vertical field MRI may not be playing a particularly
significant role in setting the saturation level in these simulations. 

\begin{figure}
\begin{center}
\subfigure{
\includegraphics[width=6.25in,height=3.125in,angle=0]{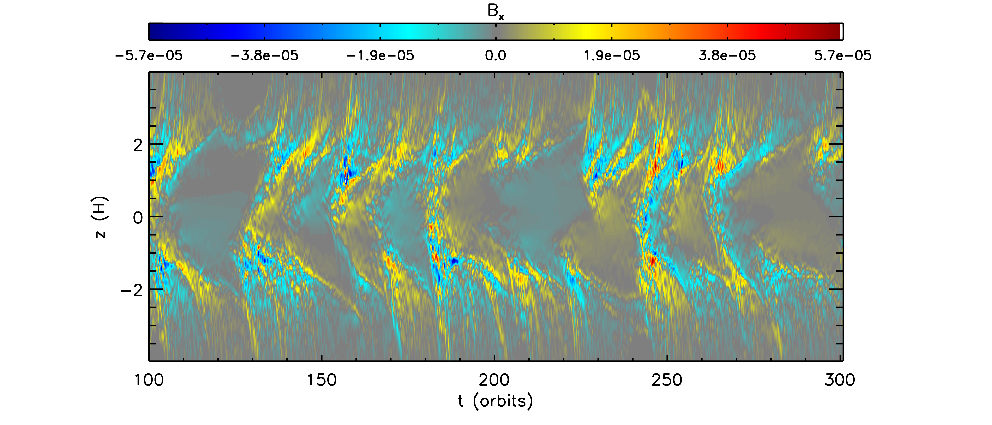}
}
\subfigure{
\includegraphics[width=6.25in,height=3.125in,angle=0]{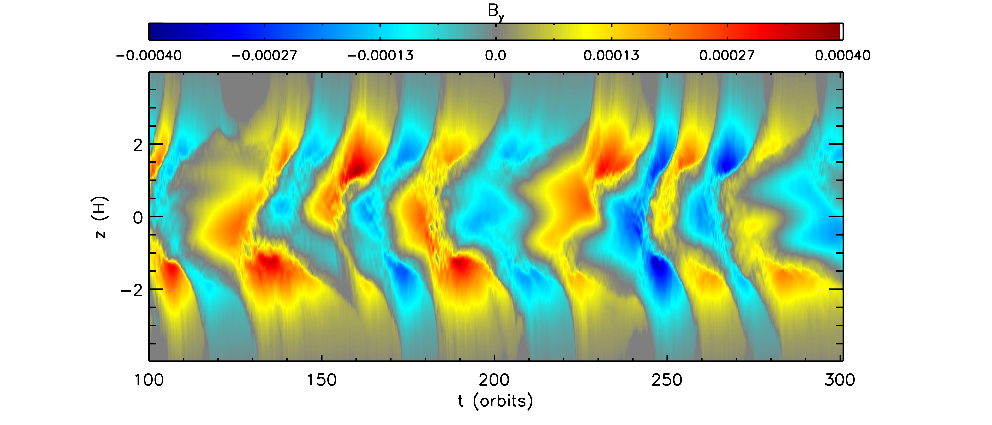}
}
\end{center}
\caption{  
Space-time diagram of the horizontally averaged $B_x$ (top panel) and $B_y$ (bottom panel) for the first 200 orbits of 32Rm800Pm0.5.
The turbulence initially decays, leaving a net $B_x$ within the mid-plane region, which shears into toroidal field.  This appears to
eventually reenergize the MRI, but the large resistivity quickly quenches the turbulence again.
}
\label{32re1600pm0.5_sttz}
\end{figure}

If this high-low variability is indeed a physical effect, what is its
origin?  First, consider the space-time diagram of the horizontally
averaged $B_x$ and $B_y$ for several simulations.
Figure~\ref{32re1600pm0.5_sttz} shows the first 200 orbits of
32Rm800Pm0.5.  The turbulence level decreases dramatically
from the beginning (see also Fig.~\ref{strat_hist}).  This is not too
surprising considering that the same $Rm$, $\pr$ values cause rapid decay
of the turbulence in the unstratified case; the resistivity is large
enough to damp the MRI.  The space-time plots show that after this decay,
there is a residual magnetic field left within the mid-plane region of
the disk.  In particular, within $|z| \lesssim 0.5 H$, there is a net
horizontally averaged $B_x < 0$ and $B_y < 0$ around $t = 110$ orbits.
The average $B_x$ within this region remains constant for awhile, and
$B_y$ increases due to the shear of $B_x$, eventually flipping to $B_y
> 0$.  By $t \sim 130$ orbits, the turbulence has re-emerged, and the
average $B_y$ rises to larger $|z|$.  The resistivity then kills off
the MRI again, leading to another period of $B_x$ shearing into $B_y$
before the next period of increased turbulence.

Model 32Rm3200Pm0.5 also experiences alternating high and low turbulence
states, but it is not immediately clear why the turbulence within the
mid-plane region should decay since the unstratified run with the same
dissipation terms has sustained turbulence.  As noted earlier and in
\cite{simon09b}, the critical $Rm$ value below which the turbulence decays
in unstratified shearing boxes is $\sim 1000$, but here $Rm = 3200$.
The major difference between the stratified and unstratified simulations
is that with vertical gravity, the net magnetic field within a localized
region of the domain can change due to buoyancy, whereas with unstratified
boxes the net toroidal field remains constant.  Loss of flux can raise
the critical $Rm$ value (numerical resolution is also
likely to have an effect); indeed, the unstratified simulations initialized with
$\beta=1000$ and $\beta=10000$ fields have demonstrated that the critical
$Rm$ depends upon the background field strength.  The loss of net flux
in a stratified box could similarly change the critical $Rm$ value.

In this simulation, the average toroidal field within the mid-plane region,
$\langle B_y\rangle$, oscillates around zero with a period of 10 orbits.
Thus, every 10 orbits or so, $\langle B_y\rangle$ is conceivably weak
enough for resistivity to kill the turbulence.  However, the turbulence remains
for many of these 10 orbit periods.  Furthermore, averaging $B_y$ within
some vertical distance from the mid-plane erases information about the
field structure there; e.g., $\langle B_y \rangle$ might be small but
there could still be strong toroidal fields of opposite polarity close to
$z = 0$.  The point is, one cannot necessarily expect the turbulence to
decay away strictly whenever $\langle B_y\rangle$ drops below a certain
(small) value.

The $\langle B_y\rangle$ oscillation amplitude
appears to be modulated by a longer timescale, more on the order of
$\sim 100$ orbits (see, e.g., Fig.~\ref{ideal_by}).  This behavior
is present in all simulations with and without physical dissipation.
This is also roughly the same timescale over which
the system switches between low and high states in the $Rm = 3200$
simulations.  Comparing the time evolution of $\langle B_y\rangle$
for $|z| \le 0.5 H$ with the evolution of the total stress shows that
the minima in the oscillation amplitude are generally correlated with
the decay of mid-plane turbulence.  One exception is near 200 orbits
in 32Rm3200Pm0.5 in which $\langle B_y\rangle$ becomes rather small,
but the turbulence remains active, though relatively weak compared
to the fully active state.  This correlation implies that if the mean
toroidal field near the mid-plane remains sufficiently small for some
time, resistivity can overwhelm the MRI and cause decay.

Evidently, there exists a critical $Rm$
below which the turbulence experiences long-timescale variability;
this critical value is $Rm < 6000$. We carry out two more stratified
simulations with $Rm \sim 6000$ but different $\pr$ values in order to
further test this hypothesis.  The first simulation is 32Rm6400Pm8; thus,
$Rm = 6400$, and $\pr$ is relatively large.  The mid-plane turbulence is
sustained over a long time, nearly 330 orbits, without any sign of decay.
The dissipation coefficients of the second simulation, 32Rm6250Pm1, are
chosen to match the relatively high $Rm$, low $\pr$ simulation that decays
in the zero net flux shearing box \cite[see][]{fromang07b,simon09b}.
This simulation also remains in the high state for nearly 330 orbits.

\begin{figure}[p]
\begin{minipage}[!ht]{8cm}
\begin{center}
\includegraphics[angle=0,width=1.25\textwidth,height=2.5in]{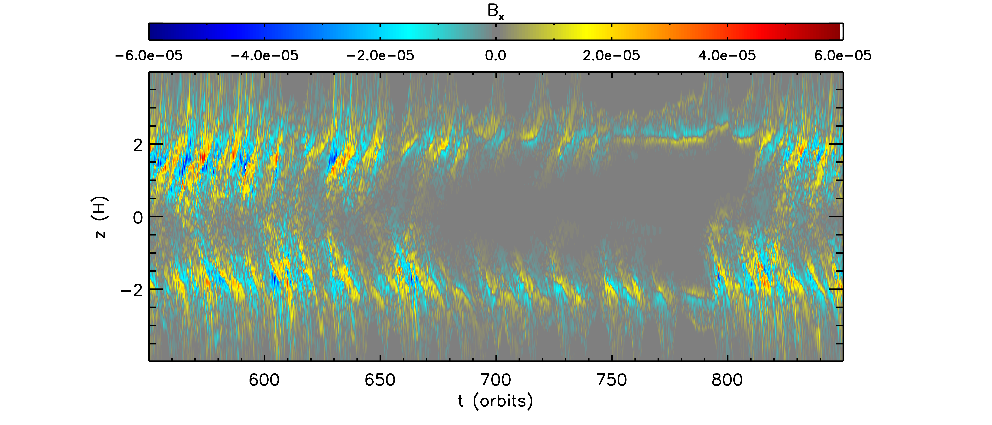}
\end{center}
\end{minipage}
\hfill
\begin{minipage}[!ht]{8cm}
\begin{center}
\includegraphics[angle=0, width=0.75\textwidth,height=2.in]{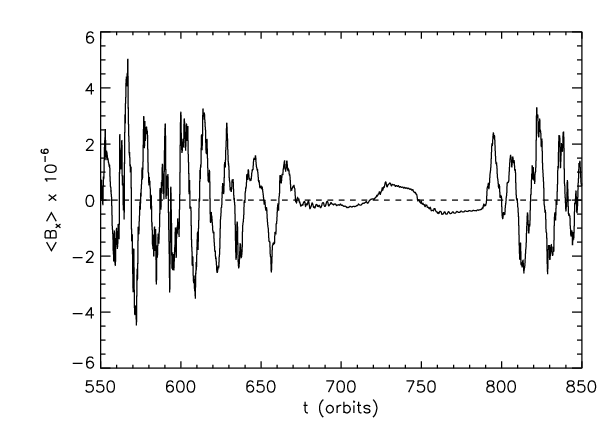}
\end{center}
\end{minipage}
\hfill
\begin{minipage}[!ht]{8cm}
\begin{center}
\includegraphics[angle=0, width=1.25\textwidth,height=2.5in]{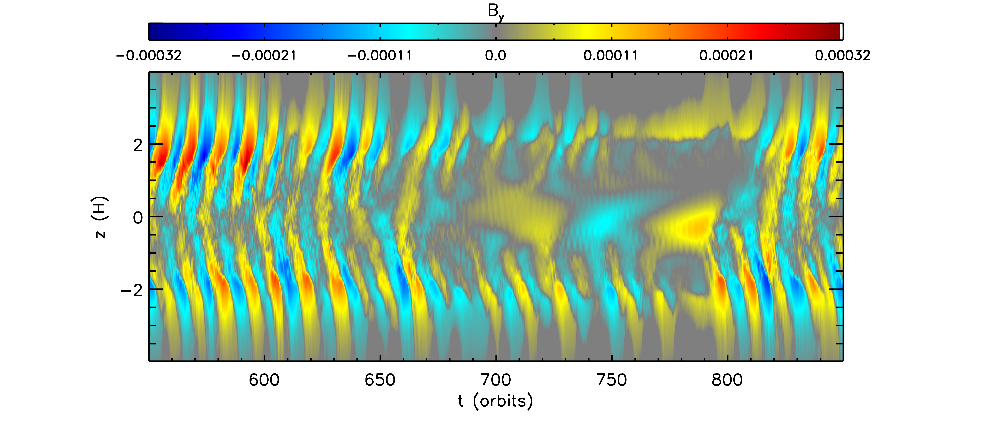}
\end{center}
\end{minipage}
\hfill
\begin{minipage}[!ht]{8cm}
\begin{center}
\includegraphics[angle=0, width=0.75\textwidth,height=2.in]{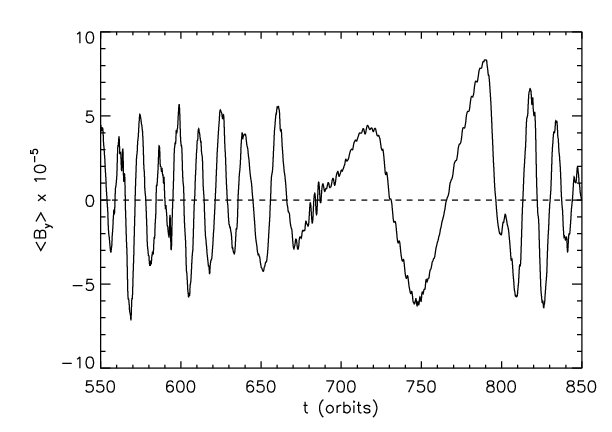}
\end{center}
\end{minipage}
\caption{Space-time diagram of the horizontally averaged $B_x$ (top left) and $B_y$ (bottom left) for a 300 orbit period in 32Rm3200Pm0.5, and 
the average of $B_x$ (top right) and $B_y$ (bottom right) over all $x$ and $y$ and for $|z| \le 0.5 H$ 
as a function of time in orbits for the same 300 orbit period.  During the period of
no MRI turbulence, a net radial field still exists within the mid-plane region.  This field appears to flip signs occasionally, leading to
corresponding flips in $B_y$ due to shear.
}
\label{32re6400pm0.5_sttz}
\end{figure}

If the dissipation is large enough to cause MRI-driven turbulence to
decay, what leads to its reactivation after $\gtrsim 100$ orbits of time?
Figure~\ref{32re6400pm0.5_sttz} shows the space-time plot of $B_x$ and
$B_y$ for a 300 orbit period in 32Rm3200Pm0.5 during which the mid-plane
turbulence dies out and eventually returns.  For clarity, we also plot
the volume-averaged horizontal field, $\langle B_x\rangle$ and $\langle
B_y\rangle$, where the average is done for all $x$ and $y$ and for $|z|
\le 0.5 H$.  During the low state, there is a small net radial field that
remains in the mid-plane region and generates $B_y$ through shear.
The sign of the net $\langle B_x \rangle$ reverses and with it the sign
of $\partial \langle B_y \rangle / \partial t$.  The last such switch in
the low state occurs around 750 orbits after which $\langle B_y\rangle$
continues to grow up until 790 orbits.  At this point, the 10 orbit
period oscillations resume.

Figures~\ref{32re1600pm0.5_sttz} and \ref{32re6400pm0.5_sttz} suggest
that it is the growth of $B_y$ due to shear that periodically reactivates
the MRI in the mid-plane.  Indeed, during the low states, the poloidal
fields are sufficiently weak that the most unstable wavelengths of the
radial and vertical field MRI are very under-resolved; $Q_x \lesssim
1$ and $Q_z \lesssim 1$ for both 32Rm800Pm0.5 and 32Rm3200Pm0.5.  ($Q$
was calculated as a function of time using equations~(\ref{alfven_q})
and (\ref{q_resolve}).)  Considering the toroidal field, however, we find
that $Q_y$ is well above the marginal resolution limit when the mid-plane
turbulence starts to decay.  In the low states of 32Rm800Pm0.5, $Q_y \sim
10-20$ occasionally dropping to $Q_y = 6$.  The typical $Q_y$ values
for the low states of 32Rm3200Pm0.5 are similar but somewhat smaller.
Of course when studying the behavior of the MRI in a situation where the
$Q$ values are small, one is in a regime where numerical resolution is
likely to matter. The specific behavior of the disk in the low state
might be different at higher resolution, but shearing of radial into
toroidal field is itself not so dependent on resolution.  Thus, $Q$
may well play a role in determining when the MRI is reactivated, but
not {\it how} it is reactivated.

\begin{figure}
\includegraphics*[scale=0.7,angle=0]{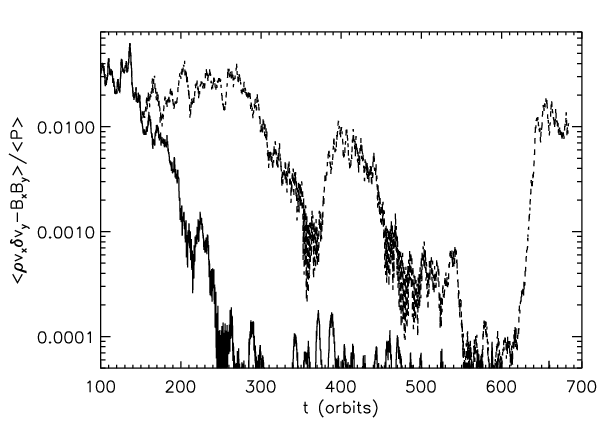}
\caption{Volume-averaged total stress normalized by the volume-averaged gas pressure
as a function of time in orbits in 32Rm3200Pm2 (solid line) and 32Rm3200Pm2\_By+ (dashed line).  The volume
average is done over all $x$ and $y$ and for $|z| \le 2 H$.  The run in which a net $B_y$ is added into the
mid-plane region (dashed line) remains in the high state at first and then exhibits variability between
low and high states.
}
\label{re1600pm2_hist}
\end{figure}

As a further test, we carry out two additional experiments.  First, we take
the state of the gas in 32Rm3200Pm2 at $t = 150$ orbits when
the stress levels are decreasing and the average of $B_y$ within $|z|
\le 0.5 H$, $\langle B_y\rangle = 5.9 \times 10^{-6}$, is relatively
small compared to the oscillation amplitude of $\langle B_y\rangle$
in the high state, which is $\sim 5 \times 10^{-5}$.  We restart this
simulation and add a net $B_y = 8.9 \times 10^{-5}$ into the region
$|z| \le 0.5 H$, which corresponds to a toroidal $\beta \approx 126$
(using $\beta$ defined with the initial mid-plane gas pressure $P_o$).
This run is 32Rm3200Pm2\_By+ (see Table~\ref{tbl:strat_sims}).
Figure~\ref{re1600pm2_hist} shows the subsequent evolution of the
stress along with the stress evolution of 32Rm3200Pm2.  Not only does
the mid-plane turbulence return, but the system undergoes episodic
transitions between low and high states on $\sim 100$ orbit time scales,
as in 32Rm3200Pm0.5.

\begin{figure}
\begin{center}
\subfigure{
\includegraphics[width=6.25in,height=3.125in,angle=0]{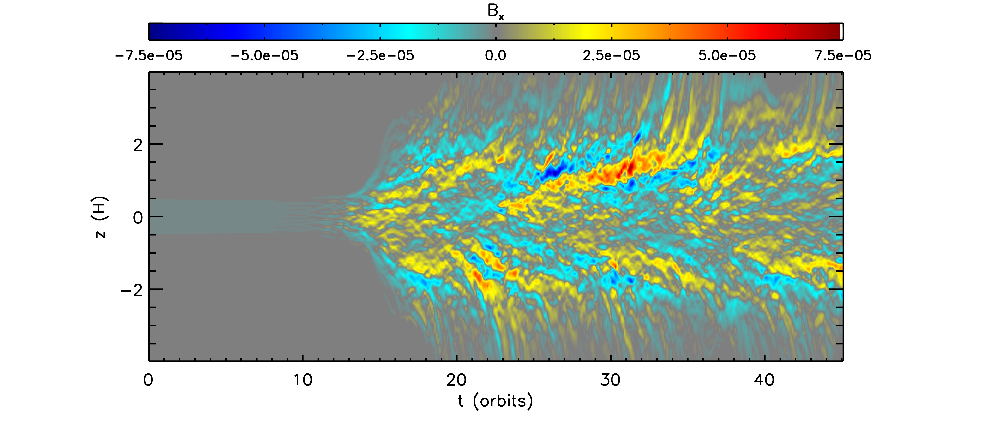}
}
\subfigure{
\includegraphics[width=6.25in,height=3.125in,angle=0]{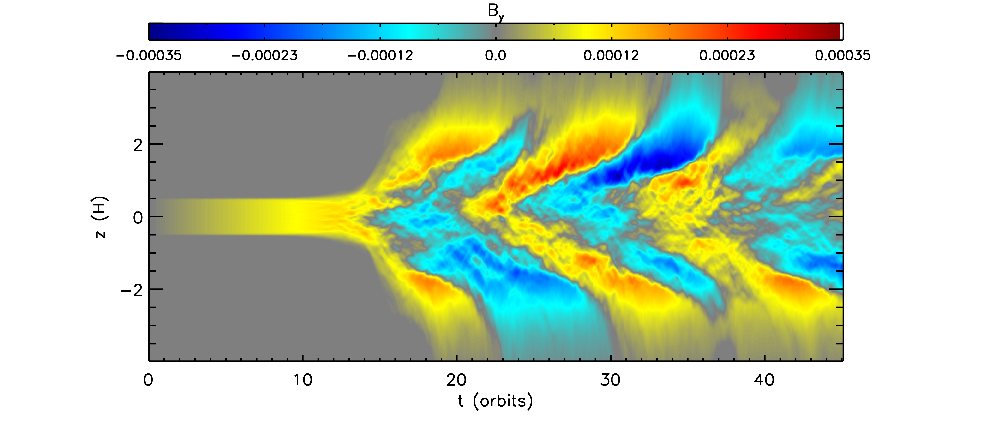}
}
\end{center}
\caption{  
Space-time diagram of the horizontally averaged $B_x$ (top panel) and $B_y$ (bottom panel) for 32ShearBx.  The uniform
radial field that is present initially shears into toroidal field, which eventually becomes strong enough to launch the MRI.
}
\label{bxshear_sttz}
\end{figure}

In a second experiment, we initialize a stratified shearing box with all
the same parameters as in 32Num but with an initially very weak radial
magnetic field.  Specifically, for $|z| \le 0.5 H$, $B_x = -\sqrt{2
P_o/\beta_x}$ where $\beta_x = 10^6$.  This field strength is very
under-resolved; the $Q_x$ value is 0.2, and the radial field MRI will
not be significant.  Figure~\ref{bxshear_sttz} shows the space-time
diagrams of horizontally averaged $B_x$ and $B_y$. The weak radial
field is sheared into toroidal field that grows linearly in time until
it reaches a sufficient strength to activate the MRI.  The onset of MRI
turbulence is rapid, and once the MRI sets in, the subsequent behavior is
very similar to the other vertically stratified MRI simulations; there are
rising magnetic field structures, dominated by the toroidal component,
and the period of oscillations in the mean field is $\sim 10$ orbits.
Note that this experiment is not an exact imitation of the low state in
our simulations; there is no MRI activity near $|z| \sim 2 H$ initially.
However, this calculation demonstrates that shear amplification of a weak radial field
can eventually lead to turbulence and move the system to the high state.

All of the simulations in which turbulence in the mid-plane sets in
after a period of quiescence show the presence of a net radial field in
the mid-plane during the low state.  32Rm3200Pm2, however, is the only
simulation that does not show the re-emergence of the mid-plane MRI, despite over
1000 orbits of integration.  An examination of the mid-plane region (up
to $|z| \lesssim H$) in the low state of this run shows that the residual
radial field is weaker than in the low states of the other simulations.
If this radial field would remain constant in time, the toroidal field
would continually strengthen to the point of reactivating the MRI.
However, $\langle B_x\rangle$ continues to change sign even in the
absence of turbulence, though with a period of many hundreds of orbits.
That is, $\langle B_x\rangle$ oscillates about zero but with a very
small amplitude.  $\langle B_y\rangle$ similarly oscillates around zero
and never reaches a sufficient amplitude to reactivate the MRI, and the
simulation remains in the low state.

In fact, $\langle B_x\rangle$ oscillates about zero in
all of our simulations, even in the low states (see e.g.,
Fig.~\ref{32re6400pm0.5_sttz}). The various space-time diagrams suggest
horizontal field migrates toward the mid-plane region from near $|z| \sim
2 H$ where MRI activity is on-going.  The oscillations in the mid-plane
field components are a direct result of the oscillations previously
observed in vertically stratified MRI turbulence.  It is not entirely
clear, however, what causes the field to migrate to the mid-plane.
Does it diffuse there or is it carried downward by some sort
of large scale flow? The shearing box calculations of \cite{turner07} and \cite{turner08},
which include Ohmic resistivity via a treatment of ionization chemistry, suggest 
that turbulent diffusion from the active layers is responsible for the field migration.
While our simulations include a simpler prescription
for resistivity, it is conceivable that a similar process is at work here.  
These issues will be addressed in future work.

To summarize the results for transitions between high and low states,
we find that for sufficiently small $Rm$, MRI turbulence within the
mid-plane region can decay away, in part because of loss of net flux
due to buoyancy.  Residual net radial field remains and subsequently
creates toroidal field from shear.  Once this toroidal field reaches a
sufficiently large amplitude, the MRI is reactivated, and the turbulence
is sustained for some duration until it decays again, repeating the
pattern. In other words, the $\alpha$--$\Omega$ dynamo, seen in
stratified simulations without resistivity, continues to operate in
resistive simulations where the mid-plane MRI is suppressed.  The dynamo is key
to reactivation of MRI-driven turbulence.

This behavior appears to be independent of the $\pr$ values we have
used, except for near $Rm = 3200$; 32Rm3200Pm4 remains in the high state,
whereas 32Rm3200Pm2, 32Rm3200Pm2\_By+, and 32Rm3200Pm0.5 do not.  However,
in the higher resolution runs, 64Rm3200Pm4 and 64Rm3200Pm0.5 show decay
but 64Rm3200Pm2 appears to be sustained (though again, these simulations
were not integrated very far).  While $\pr$ may play a role here, the
line between sustained and highly variable stress levels is unlikely to
be hard and fast, and many factors probably contribute to the nature of
the turbulence near this $Rm$ value.

Comparing our results with those of \cite{davis10}, we note that the
largest $Rm$ used in their simulations was $Rm = 3200$, consistent
with the largest critical $Rm$ observed in our simulations.  Secondly,
an examination of the space-time data from their simulation with $Rm =
1600$, $\pr = 2$ (kindly provided by the authors) shows the same behavior
as we have observed here; a net radial field remains within the mid-plane
region after decay, shearing into toroidal field, from which the MRI is
reactivated within this region.

 Interestingly, \cite{turner07} briefly notes that the
toroidal field in the dead zone of some of their highly resistive simulations can occasionally become strong enough to
reactivate turbulence there, after which the MRI is rapidly switched off again due to Ohmic dissipation. These 
simulations are consistent with our highest resistivity ($Rm = 800$) simulations, in which turbulence occurs
in short outbursts rather than through a transition to a long-lived high state, as is the case at lower resistivities.

\begin{figure}
\includegraphics*[scale=0.6,angle=270]{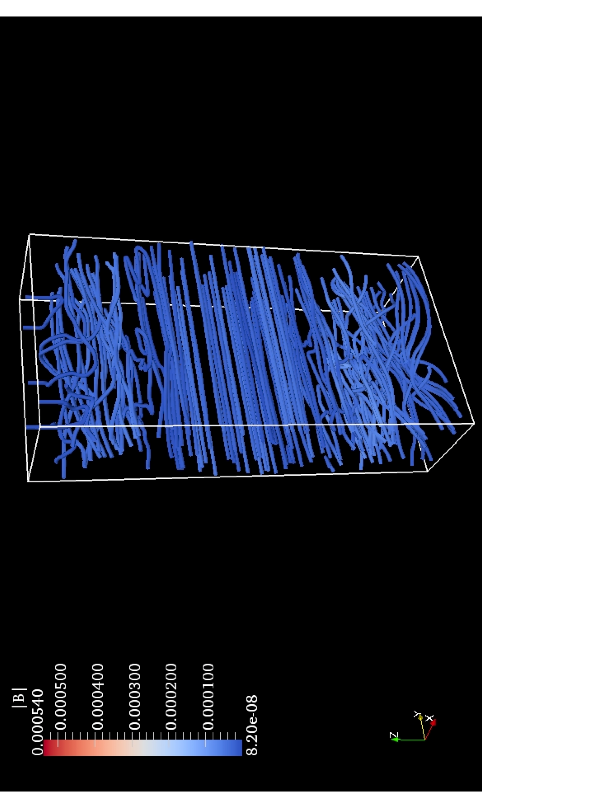}
\caption{  
Magnetic field structure at $t = 550$ orbits in 32Rm3200Pm2\_By+, produced
via a stream line integration.  The field strength (in code units) is displayed
via color and not the density of the field lines.  The color scale is the same as that in Fig.~\ref{ideal_field_lines} for
comparison.  This snapshot corresponds to the low state. The magnetic field is almost
completely toroidal near the mid-plane but has a tangled structure in the upper $|z|$ regions, more reminiscent of
the active state. 
}
\label{re1600pm2_by+_field_lines}
\end{figure}

Lastly, we examine the topology of the magnetic field in the shearing box
during the low state. Figure~\ref{re1600pm2_by+_field_lines} shows the
equivalent information as Fig.~\ref{ideal_field_lines}, but for orbit
550 of 32Rm3200Pm2\_By+.  For $|z| \gtrsim 2 H$, the field remains mainly
toroidal but with noticeable excursions from being purely toroidal, resembling the field
structure in this region during the high state. Within $2 H$, the field
is almost completely toroidal, and any small poloidal field
present within this region is not visible in this image.  We also examined
the azimuthally averaged poloidal field structure in several snapshots
of this run.  We found that the structure of the field was different,
depending on which snapshot we examined.  At some times, the poloidal
field within $2 H$ is almost completely radial, with very little vertical
field. At other times, the vertical and radial fields are comparable in
size such that the field takes on a more loop-like structure.

\subsection{The Prandtl Number Effect on Sustained Turbulence}

\begin{figure}[p]
\begin{minipage}[!ht]{8cm}
\begin{center}
\includegraphics[width=1\textwidth,angle=0]{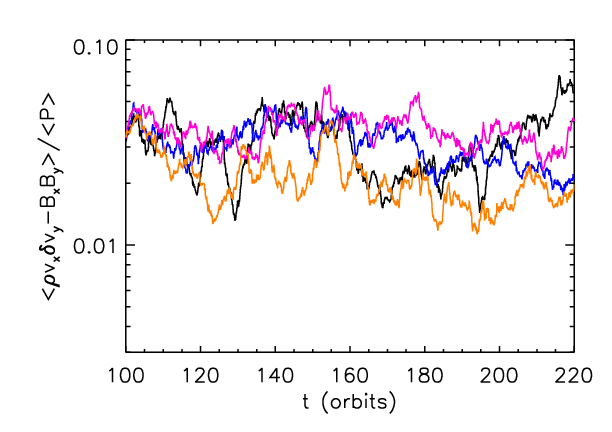}
\end{center}
\end{minipage}
\begin{minipage}[!ht]{8cm}
\begin{center}
\includegraphics[width=1\textwidth,angle=0]{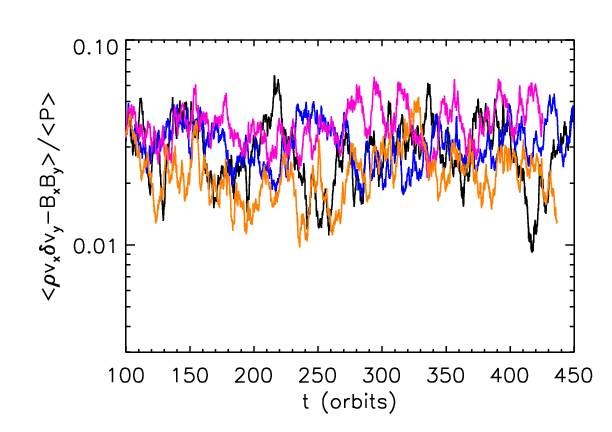}
\end{center}
\end{minipage}
\caption{
Volume-averaged total stress normalized by the volume-averaged gas pressure
as a function of time in orbits in the lower resolution, vertically stratified simulations where turbulence remains sustained. The 
volume-average is done for all $x$ and $y$ and for $|z| \le 2 H$. The left plot is the first 120 orbits of the evolution,
whereas the right plot is 350 orbits of the evolution. The black line corresponds to 
$Rm = 3200$ and $\pr = 4$, dark blue is $Rm = 6400$ and $\pr = 4$, magenta is $Rm = 6400$ and $\pr = 8$, and
brown is $Rm = 6250$ and $\pr = 1$.  The vertical axis has been chosen to match that of Fig.~\ref{ft_unstrat_hist}
for comparison.  While the stress levels generally increase with $\pr$, there is significant overlap between the different
curves at different times.
}
\label{sus_hist}
\end{figure}

In this section, we investigate how $\pr$ affects angular momentum
transport in simulations that do not exhibit the high-low variability.
Figure~\ref{sus_hist} shows the volume-averaged stress levels for these
simulations normalized by the gas pressure.  The volume average is done
for all $x$ and $y$ and for $|z| \le 2 H$.  The figure shows
a general increase in the turbulence levels with $\pr$, but there
is also significant time variability in the stress.  As a result,
the curves overlap at certain times, much more so than for unstratified
turbulence (see Fig.~\ref{ft_unstrat_hist}).

\begin{figure}[p]
\begin{minipage}[!ht]{8cm}
\begin{center}
\includegraphics[width=1\textwidth,angle=0]{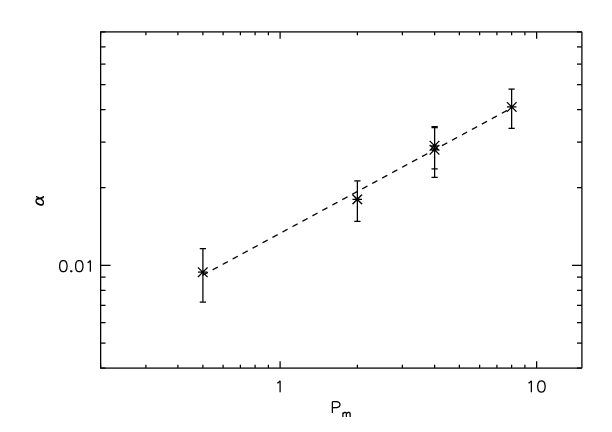}
\end{center}
\end{minipage}
\begin{minipage}[!ht]{8cm}
\begin{center}
\includegraphics[width=1\textwidth,angle=0]{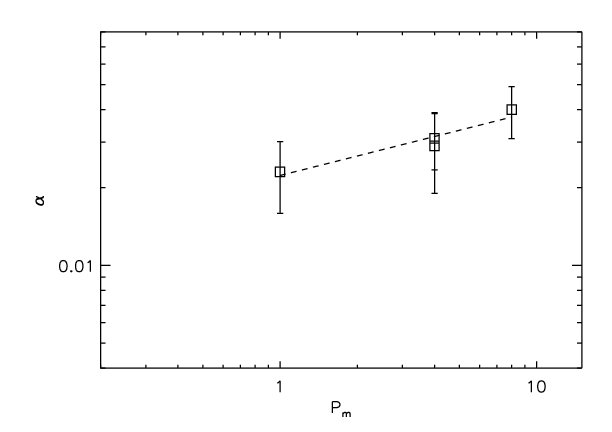}
\end{center}
\end{minipage}
\caption{
Time- and volume-averaged stress parameter $\alpha$ as a
function of $\pr$ in the unstratified FT simulations (left plot) and the stratified
simulations (right plot); $\alpha~\equiv~\left\langle\langle
\rho v_x\delta v_y - B_xB_y \rangle\rangle/\langle\langle P\rangle\right\rangle$.  The average is calculated over the entire
domain (all $x$ and $y$ and for $|z| \le 2 H$) and from 120 (150) orbits to the end of the simulation for the unstratified (stratified) runs.
  The dashed lines are linear fits to the data in log-log space,
and the error bars denote one standard deviation about the temporal average of the numerator in $\alpha$. 
Both cases show a clear $\pr$ dependence.  However, in the stratified runs, this dependence is less steep, and there
is considerable temporal variability.
}
\label{alpha_pm}
\end{figure}

Figure~\ref{alpha_pm} displays the time-averaged $\alpha$ values as
a function of $\pr$ for the unstratified (left panel) and stratified
simulations (right panel).  The time average for the unstratified
simulations is the same as in Fig.~\ref{ft_res}, from orbit 120
to the end of the calculation, and for the stratified simulations,
this average is done from orbit 150 until the end of the simulation.
The error bars denote one standard deviation about the temporal
average of the numerator in $\alpha$.\footnote{The fluctuations in
the volume-averaged pressure are very small and do not contribute
significantly to the variability.}  While there is a clear dependence
of $\alpha$ on $\pr$ in the stratified simulations,  it is not as steep
as in the unstratified simulations.  Taking a linear fit in log-log
space, and assuming $\alpha \propto \pr^\delta$,  $\delta = 0.54$ for
the unstratified runs (from \S~\ref{unstrat_sims}), and $\delta = 0.25$
for the stratified calculations.  In addition, the time variability is
significantly larger in the stratified runs than in the unstratified case.

\begin{figure}[p]
\begin{minipage}[!ht]{8cm}
\begin{center}
\includegraphics[width=1\textwidth,angle=0]{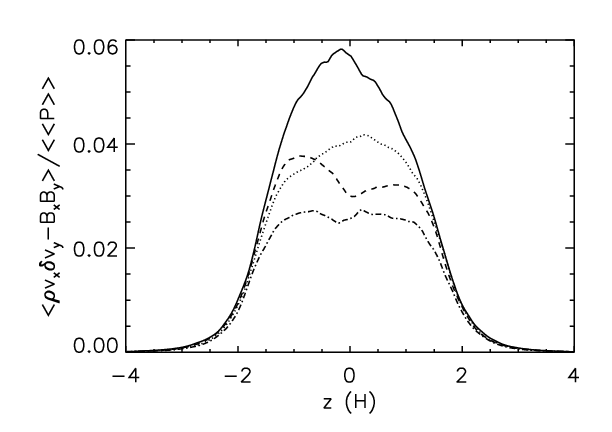}
\end{center}
\end{minipage}
\begin{minipage}[!ht]{8cm}
\begin{center}
\includegraphics[width=1\textwidth,angle=0]{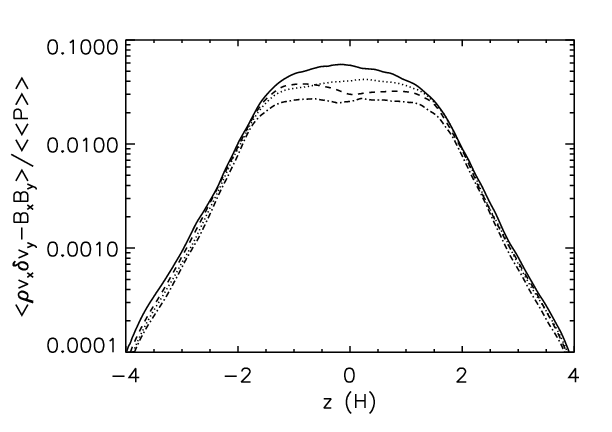}
\end{center}
\end{minipage}
\caption{
Time- and horizontally-averaged total stress as a function of $z$ on a linear (left) and logarithmic (right)
vertical scale. The stress is normalized by the time- and volume-averaged gas pressure, where the volume average is done for
all $x$ and $y$ and for $|z| \le 2 H$.  The time average is done from orbit 150 until the end of each simulation.  The solid line
corresponds to 32Rm6400Pm8, the dashed line is 32Rm3200Pm4, the dotted line is 32Rm6400Pm4, and the dot-dashed line is
32Rm6250Pm1.  The stress appears to increase with $\pr$ for nearly all $z$, and for all $\pr$, there is a sharp decrease
in the stress for $|z| \gtrsim 1.5 H$.
}
\label{stress_z}
\end{figure}

Figure~\ref{stress_z} is the vertical profile of the time- and
horizontally-averaged total stress for the sustained
turbulence runs.  The time average was done from orbit 150 to the end of
each simulation.  Increasing $\pr$ increases the stress at nearly all $z$,
and in all cases the stress drops off dramatically above $|z| \sim 1.5 H$,
consistent with the baseline simulations.  Run 32Rm6400Pm8 appears to
have a sharper peak in the stress profile near $z = 0$, compared to a
flatter stress profile for $|z| \lesssim 1.5 H$ in the other simulations.
Using different temporal averaging windows for the averaged stress
profiles produces the same general result; the stress increases with
$\pr$ and 32Rm6400Pm8 has a sharper peak near the mid-plane.  In some
cases, however, the stress does not necessarily increase monotonically
with $\pr$ at $|z| \gtrsim 1.5 H$.  Finally, we examined the vertical
profile of the same quantities as in Fig.~\ref{z_profile_ideal}.
We found that these profiles in the sustained turbulence simulations
are all very similar to each other and to 32Num; the general vertical
structure of the disk does not appear to be sensitive to $\pr$.

In summary, where the MRI is operating within a stratified disk, 
increasing $\pr$ leads to an increase in stress, consistent with the
behavior seen in unstratified simulations.

% SECTION 5

\section{Summary and Discussion}
\label{discussion}

We have carried out a series of shearing box simulations with
the \textit{Athena} code to study MRI-driven turbulence with both
vertical gravity and physical dissipation.  Until the recent work of
\cite{davis10}, the role of physical dissipation in setting the level
of angular momentum transport had been mainly studied in the context
of unstratified simulations.  As \cite{davis10} has shown, however,
stratified simulations reveal new behaviors.  Turbulence that decays
in unstratified simulations is sustained in the presence of vertical
gravity and with large amplitude, $\sim 100$
orbit fluctuations in the stress in some cases.

Our primary goal in this study has been a deeper investigation into
the roles of viscosity and resistivity in MRI-driven turbulence using simulations
that more accurately resemble astrophysical disk
systems.  To this end, we have implemented more realistic, outflow
boundary conditions in the vertical direction (\cite{davis10} use
vertically periodic boundaries), and
we have run the simulations for significantly longer integration times
than is usual for shearing box calculations, in order to study long timescale effects.
From these simulations we observe the following:
1) MRI turbulence is largely confined within $2~H$ of the equator.
Above that height, the magnetic field becomes completely buoyantly unstable.
2) The mean field within the MRI-active portion of the disk evolves in
a manner consistent with an $\alpha$--$\Omega$ dynamo.
3) Stratified disks show an increase in stress with increasing $\pr$
as previously seen in unstratified shearing boxes.
4) For $Rm$ below some critical value, MRI turbulence in the mid-plane
can be quenched putting the disk into a ``low'' state.
5) While in the low state, the continued action of the dynamo can raise
the toroidal field to a level at which the mid-plane MRI is reactivated,
switching the disk back to a ``high'' state.

The first two of these results are consistent with previous stratified
shearing box simulations and do not appear to be altered by the inclusion
of physical dissipation.  For $|z| \lesssim 2 H$ in sustained turbulence
simulations, the time- and horizontally-averaged turbulent energies and
stresses are roughly constant with height, the magnetic field is only
marginally stable to buoyancy, and $\beta > 1$.   The predominantly
toroidal field rises slowly away from the midplane.  Above this height,
the turbulence is significantly weaker, $\beta < 1$, and the field is completely
buoyantly unstable, rising at a faster rate than for $|z| \lesssim 2 H$.
These results, which are consistent with the ZEUS-based results of
\cite{guan10} and \cite{shi10}, suggest that there are two separate vertical regions
in these disks.

As a practical matter, this result suggests that to capture the behavior
of the vertically stratified MRI, one need not go much beyond $\pm 2 H$
from the mid-plane.  This is confirmed by a few additional simulations
in which we extended the outer boundary to $6 H$ from the mid-plane
instead of $4 H$.  We found no difference in the vertical structure of
the disk, the volume averaged stress levels, or the temporal variability
of the system.

Consistent with previous stratified shearing box simulations
\citep[e.g.,][]{stone96a,hirose06,guan10,shi10,gressel10,davis10} and
 global simulations \cite[]{fromang06b,dzyurkevich10,oneill10},
we found a strong 10 orbital period variability in the mean magnetic
field within the mid-plane region.  This variability is characterized
by the buoyant rise of predominantly toroidal field from the mid-plane
region, after which a field of opposite sign takes its place at the
mid-plane.  The evolution of this toroidal field can be well-modeled by
a simple $\alpha$--$\Omega$ mean field dynamo where the shear of
radial field into toroidal field and the subsequent buoyant removal of
toroidal field from the mid-plane dominate the evolution.

In addition to the dynamo period of roughly 10 orbits, we also observe
a longer $\sim 100$ orbit timescale variability.
This behavior has not previously been reported, and while we do not know
its origin, there is evidence that it plays a role in another
novel effect resulting from vertical gravity: high and low states.
For disks that are sufficiently resistive, i.e., have low enough $Rm$, MRI
turbulence decays near the mid-plane leading to a quiescent period that is
eventually followed by regrowth of the MRI in this region.  The period of
this variability is on a similar timescale to the long-period variability
seen in all the stratified simulations.  The transition to a low state
occurs when the mean toroidal field in the mid-plane region is reduced
below some critical value, presumably by buoyancy.  Resistivity dominates
over the MRI and the turbulence decays.  An extremely weak mean radial
field remains however.  Although this mean radial field itself can
vary during the low state, it creates toroidal field through shear.
Shear amplification is largely unaffected by resistivity.  Once the
toroidal field reaches a particular strength, the mid-plane MRI is re-energized,
and the disk becomes fully turbulent again, returning to the high state.

This behavior is not particularly sensitive to $\pr$ and appears to be
the same as that reported by \cite{davis10}. Thus, it is not likely
to be related to the dynamo issue of $\pr \sim 1$ in zero net magnetic
flux shearing boxes, which \cite{davis10} specifically investigated.
Instead, it appears to be a robust behavior that emerges whenever the disk is sufficiently resistive.  The critical
$Rm$ below which the high-low variability exists is $3200 \lesssim Rm_c \lesssim 6000$.  If $Rm > Rm_c$,
the turbulence remains sustained for the dissipation parameters explored
here, and averaged stress increases with $\pr$ for all $z$, though with
a less steep dependence of $\alpha$ on $\pr$ compared to unstratified
simulations.

Because we are dealing with the decay of turbulence and the behaviors
of weak magnetic fields, the values of properties such as $Rm_c$ may
well be resolution dependent.  However, given the long evolution times probed
here, it was necessary to limit the resolution to 32 grid zones per $H$.
Before running the vertically stratified simulations, we executed a series
of unstratified calculations at this resolution to test whether or not the
MRI and the effects of physical dissipation would likely be resolved. We
found that the effect of physical dissipation on MRI saturation is
reasonably well converged at 32 zones per $H$: $\alpha$ increases with
$\pr$ at this resolution, although with a steeper dependence than at
higher resolution, and $\alpha$ is roughly constant with resolution above
32 zones per $H$ for a given set of $Re$, $Rm$, and $\pr$ values.  This in
itself is an interesting result since it shows that the higher resolutions
used in \cite{fromang07b}, \cite{lesur07}, and \cite{simon09b} may not be
necessary to capture the general effects of physical dissipation. It is
also in agreement with the recent vertically stratified shearing boxes
of \cite{flaig10} that included radiation physics; they found that only
$\sim 30$ zones per $H$ in the vertical dimension are required to gain
a reasonable representation of the turbulent saturation level.  Thus, while the exact
value of $Rm_c$ may change at higher resolution, the general effects of physical
dissipation on the vertically stratified MRI are likely captured at the resolutions we used. 

What do these results imply for the MRI and its application to
astrophysical disks?  First, it had been previously appreciated
that sufficient resistivity could cause a transition from a turbulent
(``high'') state to a non-turbulent (``low'') state.  Here we have found
that a process resembling an $\alpha$--$\Omega$ dynamo can accomplish
the reverse and re-establish the high state.  The transition occurs on
longer timescales than the $\sim 10$ orbit associated with the dynamo.
This temporal variability could have potential applications for several
types of accretion disks.  Protoplanetary disks have large regions of
low ionization gas, including a dead zone layer \cite[]{gammie96} where
resistivity is too high to sustain the MRI.  Dwarf nova disks also contain
regions of partial ionization, and it is intriguing that the $Rm$ values
in these systems are on the same order as the critical $Rm \sim 10^3$
for the decay/regrowth behavior \cite[]{gammie98}.  Even some regions
of AGN disks may have moderately high resistivity, though typical $Rm$
values are probably larger than those in dwarf nova systems because of
the larger disk scale height \cite[]{menou01}.

It is tempting to associate the peaks and dips of turbulent activity in
our simulations with the outbursts and variability observed in these
systems.  Of course, there remains much work before such a connection
can be substantiated.  In particular, more realistic simulations
will have $\eta$ (and $\nu$) that depend on temperature and density,
rather than being constant throughout the disk.  Furthermore, the
influence of other non-ideal MHD effects on the MRI needs more work.
The Hall effect is often times just as important as Ohmic resistivity
in astrophysical environments \cite[]{wardle99,balbus01,balbus03}, and
while simulations including both Hall and Ohmic terms have been carried
out by \cite{sano02a} and \cite{sano02b}, there remains more parameter
space to explore and physics to include.  Lastly, we note that if $\eta$
is so large that no MRI modes are present, there would be no temporal
variability and the turbulence would be completely quenched.

In disks that have $Rm$ above the critical value, the MRI-driven turbulence
operates continuously.  Nevertheless, as previously seen in unstratified
simulations, dissipation terms can have an impact: the stress level
increases with increasing $\pr$.  This may be relevant to hot, fully
ionized disk gas such as in X-ray binaries and some regions of AGN disks where $\pr$
has a strong temperature dependence \citep{balbus08a}.  While our
simulations show that angular momentum transport increases with $\pr$,
the $Re$ and $Rm$ values of such disks are significantly larger than the
values probed here.  Whether or not the $\pr$ effect continues into the
large $Re$/$Rm$ regime remains very much an open area of research.

Finally, one particular field geometry that has not been explored here
or in most vertically stratified local simulations is that of a net
vertical field.  These simulations are quite challenging; the channel
mode dominates the solution \cite[]{miller00,latter10}, leading to
very strongly magnetized regions of the disk that can often times cause
the numerical integration techniques to fail (but see \cite{suzuki09},
in which a stable evolution was produced).

In summary, we have explored the spatial and temporal behavior of the
MRI in the presence of both vertical gravity and physical dissipation.
We find that for moderately resistive simulations, the disk cycles
between states of low and high turbulent stresses, and that orbital
shear of radial field into toroidal field is essential to both this
behavior as well as the temporal variability of sustained turbulence.
When sustained, the stress increases with $\pr$, in agreement with
unstratified simulations. Our calculations are an important stepping stone
towards more realistic simulations that include temperature-dependent
$\nu$ and $\eta$.

\acknowledgments

We thank Xiaoyue Guan, Juilan Krolik, Mitch Begelman, Phil Armitage, and Rosalba Perna for useful discussions and
suggestions regarding this work, and we are very grateful to Shane Davis,
Jim Stone and Martin Pessah for providing us with some of the data
from their paper.  This material was supported by NASA Headquarters
under the NASA Earth and Space Science Fellowship Program - Grant
NNX08AX06H; NASA grants NNX09AG02G, NNX09AB90G, and NNX09AD14G;
by the NSF under grants AST-0807471 and AST--0908869; and by a Virginia Space Grant Consortium (VSGC) fellowship.
The simulations were run on the TACC Sun Constellation Linux Cluster, Ranger, supported
by the National Science Foundation.

% APPENDIX

\appendix

\section{Numerical Methodology}

\subsection{Integration}
\label{integration}

The algorithms for solving the shearing box 
equations (\ref{continuity_eqn})--(\ref{induction_eqn}) 
and the specific implementation of those algorithms 
within the \textit{Athena} code are described in \cite{stone10}; here we
provide a brief summary of the numerical methods detailed
in that paper.

The left hand sides of the equations are solved via the standard
\textit{Athena} CTU flux-conservative algorithm \cite[]{stone08}.
The gravitational and Coriolis source terms in the momentum equation
are evolved via a combination of an unsplit method consistent with CTU
and Crank-Nicholson differencing constructed to conserve energy exactly
within epicyclic motion \cite[]{gardiner05a,stone10}.

For our simulations in which the radial size of the shearing box is a scale height, $H$,
the velocity is initialized with a background shear flow given by
\begin{equation}
\label{shear_flow}
v_y = -q \Omega x.
\end{equation}

\noindent For sufficiently large $x$ domains, this velocity can become
supersonic (orbital velocities are, in general, supersonic in accretion disks).   For large $x$ domains, the
Courant limit on the timestep can become significant.  Furthermore, the
background shear flow can lead to a systematic change in truncation error
with radial position in the box, which in turn introduces purely numerical
features in the radial density and stress profiles \cite[]{johnson08}.
Hence, for large radial domain simulations we utilize an orbital
advection scheme, which subtracts off this background shear flow and
evolves it separately from the fluctuations in the fluid quantities
\cite[]{masset00,johnson08,stone10}.

\subsection{Boundary Conditions}

In the shearing box, the $y$ boundary condition is periodic.  The $x$
boundary condition is shearing-periodic \citep{hawley95a,stone10}:
quantities are reconstructed in the ghost zones from appropriate zones in
the physical domain that have been shifted along $y$ to account for the
relative shear from one side of the box to the other.  In {\it Athena},
this reconstruction step is performed on the fluid fluxes, and then the
ghost zone fluid variables are updated via these reconstructed fluxes.
The order of this reconstruction matches the spatial reconstruction in the
physical grid, e.g., 3rd order reconstruction of the ghost zone fluxes
is done when the PPM spatial reconstruction is employed.   Furthermore,
the $y$ momentum is adjusted to account for the shear across the $x$
boundaries as fluid moves out one boundary and enters at the other. The
$y$ component of the electromotive force (EMF) is reconstructed at the
radial boundaries to ensure precise conservation of net vertical magnetic
flux \cite[]{stone10}.

In unstratified shearing box simulations the $z$ boundary is 
periodic.  This same boundary condition has been employed in
stratified boxes as well \citep[e.g.][]{davis10}.  For our simulations,
we use an outflow boundary condition instead.  
The density $\rho$ is extrapolated into the ghost
zones based upon an isothermal, hydrostatic equilibrium, using the last
physical zone as a reference. Therefore, for the upper vertical boundary,
the $\rho$ value in grid cell $k$ is

\begin{equation}
\label{density_bc}
\rho(k) = \rho(ke) {\rm exp}\left(-\frac{z(k)^2-z(ke)^2}{H^2}\right),
\end{equation}

\noindent
where $ke$ refers to
the last physical zone at the upper boundary.  An equivalent expression
holds for the lower vertical boundary. This extrapolation provides
hydrostatic support against the opposing vertical gravitational forces,
which are also applied in the ghost zones.  All velocity components,
$B_x$, and $B_y$ are copied into the ghost zones from the last physical
zone assuming a zero slope extrapolation.  If the sign of $v_z$ in the
last physical zone is such that an inward flow into the grid is present,
$v_z$ is set to zero in the ghost zones.  Finally, the ghost zone values
of $B_z$ are calculated from $B_x$ and $B_y$ to ensure that $\del \cdot
{\bm B} = 0$ within the ghost zones.

\subsection{Physical Dissipation}

Viscosity and resistivity are added via operator splitting; the
fluid variables updated from the main integration (\S~\ref{integration})
are used to calculate the viscous and resistive terms.  The viscosity
term is calculated via the divergence of the viscous stress tensor,
equation~(\ref{viscous_stress_tensor}), and the resistive term is
included as an additional EMF within the induction equation
(\ref{induction_eqn}).  This formulation allows us to
discretize the viscous and resistive terms in a flux-conservative and
constrained-transport manner, consistent with the \textit{Athena} algorithm.
Note that the resistive contribution to the $y$ EMF must also be
reconstructed at the shearing-periodic boundaries in order to preserve
$B_z$ precisely.

The addition of viscosity and resistivity places an
additional constraint on the timestep,

\begin{equation}
\label{timestep}
\Delta t = {\rm MIN}\left(\Delta t_{\rm CTU}, C_o \frac{\Delta^2}{4 \nu}, C_o \frac{\Delta^2}{4 \eta}\right),
\end{equation}

\noindent 
where $\Delta t_{\rm CTU}$ is the timestep limit from the main
integration algorithm \cite[see][]{stone08}, $C_o = 0.4$ is the CFL number, and $\Delta$ is the
minimum grid spacing, $\Delta = {\rm MIN}(\Delta x, \Delta y, \Delta z)$.
Note that in most of our simulations $\nu$ and $\eta$ are sufficiently
small that they do not limit the timestep.

\clearpage

\clearpage

\clearpage

\clearpage

\clearpage

\clearpage

\clearpage

\clearpage

\clearpage

\clearpage

\clearpage

\clearpage

\clearpage

\clearpage

\clearpage

\clearpage

\clearpage

\clearpage

\end{document}